\documentclass[preprint]{JHEP3}
\usepackage{cite}
\usepackage{amssymb,amsmath}
\usepackage{graphicx}
\usepackage{epsfig}

\graphicspath{{figures/}}

\def\Q2{\left(Q^{2}\right)}

\def\eps{\epsilon}

\def\l({\left(}
\def\r){\right)}

\def\eps{\epsilon}

\def\l{(}
\def\r{)}

\title{Photon isolation effects at NLO in $\gamma\gamma$+jet final states in 
hadronic collisions}
\author{T.\ Gehrmann$^a$, N.\ Greiner$^b$, G.\ Heinrich$^b$\\
$^a$ Institut f\"ur Theoretische Physik, Universit\"at Z\"urich,
Wintherturerstrasse 190,\\CH-8057 Z\"urich, Switzerland\\
$^b$ Max Planck Institut f\"ur Physik, F\"ohringer Ring 6, D-80805 M\"unchen, Germany}

\abstract{We present the NLO QCD corrections to $pp\to \gamma\gamma j$ production at 
hadron colliders. Our calculation includes contributions from the fragmentation of a hadronic 
jet into a highly energetic photon, and consequently allows the implementation of 
arbitrary infrared-safe photon isolation 
definitions. We compare different photon isolation criteria and perform a 
detailed study of the dependence of the 
$\gamma\gamma j$ cross section on the photon isolation parameters. 
 }
\preprint{{ZU-TH 05/13, MPP-2013-32}}

\begin{document}

\bibliographystyle{JHEP}

\allowdisplaybreaks

\section{Introduction}

Diphoton final states have played a crucial role in  the recent discovery of a new 
boson at the LHC~\cite{:2012gk,:2012gu}, and the fact that its partial decay
width  to photons seems to be enhanced as compared to the Standard Model (SM) Higgs boson 
keeps up the interest in this channel. 
Diphotons are also important in many New Physics searches~\cite{:2012afa,Aad:2012cy,CMS:2012un,:2012mx}, 
in particular the search for extra spatial dimensions or cascade decays 
of heavy new particles. 
In particular,  diphotons in combination with jets and missing energy 
occur in gauge mediated SUSY scenarios.

In what concerns the SM processes at the LHC, 
the diphotons will most certainly be accompanied by one or more high-$p_T$ jets, 
which are often suppressed by a jet veto. 
However, the presence of an extra jet offers better control on 
the separation of backgrounds and signals 
and more information on the interaction dynamics.
Therefore a precise understanding of SM processes yielding diphotons in association with jets
is mandatory.

Diphoton production has been calculated at NLO some time ago~\cite{Binoth:1999qq}, supplemented also by 
gluon initiated subprocesses beyond the leading order~\cite{Bern:2002jx} 
and soft gluon resummation~\cite{Balazs:1999yf,Balazs:2006cc}. Recently,  NNLO corrections to 
direct diphoton production also have become available~\cite{Catani:2011qz}.
Monte Carlo approaches to prompt photon production with matrix-element/parton-shower merging  also have been 
studied~\cite{Hoeche:2009xc,D'Errico:2011sd,Odaka:2012ry}. 
In \cite{Hoeche:2009xc}, the fragmentation contribution 
is generated by a combined QCD+QED shower, and photon isolation is done within the 
democratic clustering approach~\cite{Glover:1993xc,GehrmannDeRidder:1997wx,GehrmannDeRidder:1998ba}.
Refs.~\cite{D'Errico:2011sd,Odaka:2012ry} contain an NLO Monte Carlo study of diphoton production, 
where fixed cone-based photon isolation is also possible.

NLO calculations of diphoton-plus-jet production~\cite{DelDuca:2003uz}, 
photon-plus-two-jet production in QCD production~\cite{Bern:2011pa} and in vector boson fusion~\cite{Jager:2010aj}, 
and diphoton  production 
at NNLO~\cite{Catani:2011qz} were up to now only carried out based on a smooth-cone isolation and did not 
admit the implementation of  alternative isolation criteria. 

In this article, we present an NLO calculation of diphoton production in association with one jet
which also contains a component from the fragmentation of QCD partons into photons, 
thereby allowing  to study the effect of different photon isolation criteria.
In addition, we provide a public code to compute 
$\gamma\gamma$+jet final states at NLO, where 
the virtual amplitude has been pre-generated with 
the automated one-loop program {\sc GoSam}~\cite{Cullen:2011ac}. 
Section \ref{sec:frag} contains a general discussion of photon isolation criteria and fragmentation.
In Section \ref{sec:calc} technical details of the calculation are discussed, while the results are presented in  
Section \ref{sec:results}, before we conclude in Section \ref{sec:conclusion}.

\section{Photon isolation criteria and photon fragmentation}
\label{sec:frag}
Photons in hadronic final states can have multiple origins. Besides the radiation of a hard 
photon off a quark involved in the hard interaction (sometimes called 'prompt' photon), 
photons can also be emitted during the hadronization phase of a hadronic jet or can be 
the result of electromagnetic decays of unstable hadrons (these are sometimes 
called 'secondary photons'). Especially the decay 
$\pi^0\to \gamma \gamma$ of pions at high transverse momentum can mimic the 
signature of a single photon if the two photons are too closely collimated to be 
resolved individually. Only the production of prompt photons can be computed 
within perturbation theory from first principles, while the production of photons in hadronization
and hadron decays can only be modeled, thereby introducing dependence on ad-hoc parameters. 

Photons resulting from the hard interaction are usually well-separated from all hadronic jets
produced in the event, while photons from hadronization and decay processes will always be 
inside hadronic jets. To disentangle prompt from secondary photons,
one applies isolation requirements, which limit the hadronic activity around a photon 
candidate, thereby defining an ``isolated photon". A veto on all hadronic activity around the 
photon direction would result in a suppression of soft gluon radiation in part of the final state 
phase space, thereby violating infrared safety of the observables. Consequently, all 
photon isolation prescriptions must admit some 
residual amount of hadronic activity around the photon direction. 
 
By admitting some hadronic activity around a photon, one includes final state configurations 
with a final state quark radiating a highly energetic collinear photon. These configurations contain 
a collinear singularity, related to small invariant masses of the quark-photon system. Mass 
factorization in QCD relates this singularity to a redefinition of the quark-to-photon 
fragmentation function, which describes the production of a photon inside a hadronic jet. 
Like parton distributions in the proton, these fragmentation functions are non-perturbative 
objects that have to be determined from experimental observations. Their dependence on 
the resolution scale is governed by evolution equations. 

The fragmentation contribution intertwines the production of prompt and secondary
 photons. An observable with final state photons defined through isolation criteria will typically 
 receive some contribution from photon fragmentation, with the isolation criterion aiming to 
 minimize this contribution. Several isolation criteria were proposed in the literature and 
 applied in experimental studies of single photon and photon pair production. 

The cone-based isolation is most commonly used especially at hadron collider 
experiments. 
In this procedure, the photon candidate is identified (prior to the jet clustering) 
from its electromagnetic signature, 
and its momentum direction (described by 
transverse energy $E_{T,\gamma}$, rapidity $\eta_\gamma$ and polar angle $\phi_\gamma$) 
is determined. Around this momentum direction, a cone of radius 
$R$ (typically chosen around 0.4) 
in rapidity $\eta$ and polar angle $\phi$ is defined. Inside this cone, the hadronic 
transverse energy  $E_{T,{\rm cone}}$ is measured. The photon is called isolated if 
 $E_{T,{\rm cone}}$ is below a certain threshold, defined either in absolute terms, or as 
 fraction of $E_{T,\gamma}$ (typically 0.1 or below).  
 The latter criterion then means that a photon candidate is considered as isolated if  in a cone 
 defined by 
 \begin{displaymath}
(\eta-\eta^\gamma)^2+(\phi-\phi^\gamma)^2\leq R^2\;,
\end{displaymath}
the amount of hadronic energy does not exceed a pre-defined fraction of 
the photon energy: 
\begin{equation}
E_{{\rm had}}  \leq \epsilon_c \,p_{T}^{\gamma} \;.
 \label{eq:cone-iso}
 \end{equation}
 
 An alternative to the cone-based isolation is the democratic clustering 
 procedure~\cite{Glover:1993xc}. In this procedure, a photon 
 candidate is treated like any hadron in the 
 jet recombination, which can be 
 performed using any (cone-based or clustering based) infrared-safe jet 
 algorithm~\cite{Salam:2009jx}.
  After the jet recombination, jets containing photons are labeled as photon jets, which 
 are identified as isolated photons if the photon energy in the jet exceeds a predefined 
 fraction $z_{\rm cut}$ of the jet energy. Typical values of $z_{\rm cut}$ are 0.9 or above. 
 
 Both cone-based isolation and democratic clustering admit some fraction of events involving 
 collinear quark-photon systems. The theoretical predictions for cross sections 
 defined in these procedures must therefore take account of photon fragmentation contributions. 
 
 Finally, the smooth cone isolation criterion~\cite{Frixione:1998hn} varies the 
 threshold on the hadronic energy inside the isolation cone with the 
 radial distance from the photon. It is described by the cone size $R$, 
 a weight factor $n$ and 
 an isolation parameter $\epsilon$. With this criterion, one considers smaller 
 cones of radius $r_\gamma$ inside the $R$-cone and calls the photon isolated 
 if the energy in any sub-cone does not exceed
 \begin{equation}
 E_{{\rm had, max}} (r_{\gamma}) = \epsilon \,p_{T}^{\gamma} \left( \frac{1-\cos r_\gamma}
 {1-\cos R}\right)^{n}\;.
 \label{eq:frix}
 \end{equation}
 In theoretical studies, typical values 
 used for the smooth cone isolation parameters are $R=0.4$, 
 $\epsilon=0.5$, $n=1$. By construction, the smooth cone isolation 
 does not admit any hard collinear quark-photon configurations, thereby allowing a full 
 separation of 
 direct and secondary photon production, and consequently eliminating the need for 
 a photon fragmentation contribution in the theoretical prescription. 
 Despite this advantage, the smooth cone isolation was up to now 
 used in experimental studies of isolated photons
 only in a discretized approximation\,\cite{AlcarazMaestre:2012vp}. 
 Owing to finite detector resolution, an 
 implementation will only be possible to some minimal value of $r_{\gamma}$, thereby 
 leaving potentially a residual collinear contribution. 
 
 Perturbative calculations of isolated photon production must take proper account of the 
 isolation criterion used to define the observable. Besides the usual higher-order QCD effects
from unresolved partonic radiation, these calculations must also take account of 
the quark-photon collinear singularity appearing in the photon isolation procedure.
In some specific observables (inclusive photon production in 
$e^+e^-$-annihilation~\cite{Kunszt:1992np,Ackerstaff:1997nha}
and deep inelastic scattering~\cite{GehrmannDe Ridder:2006wz,Chekanov:2004wr,Chekanov:2009dq,Aaron:2007aa} 
as well as 
photon-plus-one-jet production in $e^+e^-$~\cite{Glover:1993xc,Buskulic:1995au,GehrmannDeRidder:1997wx,GehrmannDeRidder:1997gf} 
and photon-plus-no-jet production 
in deep inelastic scattering~\cite{Aaron:2007aa}), this singularity appears already at the leading 
order~\cite{Glover:1993xc,GehrmannDe Ridder:2006wz}. 
These observables are consequently most sensitive 
to the photon fragmentation function and can thus be used for its 
determination~\cite{GehrmannDeRidder:1998ba,GehrmannDeRidder:2006vn}. 

Concerning the counting of perturbative orders, some ambiguity arises when photon 
fragmentation contributions are present. The photon fragmentation function itself is 
${\cal O}(\alpha)$. Its evolution equation differs from the evolution equations for 
the fragmentation functions of hadrons~\cite{Altarelli:1977zs} by a direct term, which results in
logarithms of the resolution scale not being suppressed by one order of the strong coupling 
constant (as in the case of hadron fragmentation). Motivated by this difference, it has been 
argued that the photon fragmentation function should be assigned an inverse power of the 
strong coupling constant~\cite{Owens:1986mp,Gluck:1992zx,Bourhis:1997yu}, 
thereby shifting the relative order of 
direct and fragmentation contributions. Viewed by mass factorization counter terms 
(and infrared finiteness of the observables), the photon fragmentation function does not 
require an inverse power of the strong coupling constant. In calculations of electroweak 
corrections to jet observables~\cite{Denner:2009gx,Denner:2010ia,Denner:2009gj,Denner:2011vu} where similar photon isolation issues appear, 
the photon fragmentation function is consistently taken as ${\cal O}(\alpha)$. 

Single photon and diphoton production at hadron colliders receive contributions from 
collinear quark-photon splitting only at next-to-leading order in perturbation theory. 
Among the existing calculations of NLO corrections to 
photon pair production~\cite{Binoth:1999qq,Balazs:1999yf,Bern:2002jx} and photon-plus-jet 
production~\cite{Catani:2002ny,Belghobsi:2009hx}, 
the ones in~\cite{Binoth:1999qq,Catani:2002ny,Belghobsi:2009hx}
have implemented both cone-based and smooth cone 
isolation and assume the power counting of the photon fragmentation function 
to contain an inverse power of the strong coupling constant, which means that a multitude 
of fragmentation subprocesses  have to be (and have been) included at NLO. 
In general, these subprocesses are however 
of rather minor impact on the total result after isolation cuts, owing to the overall 
smallness of fragmentation contributions in these observables. For the purpose of our 
calculations, we will therefore use a power counting of the photon fragmentation function as
${\cal O}(\alpha)$.

\section{NLO corrections to $pp\to \gamma\gamma$+jet final states}
\label{sec:calc}

\subsection{Structure of the calculation}
The calculation of the NLO corrections to $pp\to \gamma\gamma$+jet requires the combination
of the full QCD corrections with counter terms regularizing infrared QED singularities.
In a first step we produced a code that is able to calculate the QCD corrections. For generation
of the tree level and real emission matrix elements we use MadGraph \cite{Stelzer:1994ta,Alwall:2007st},
the regularisation of infrared QCD singularities is handled by MadDipole \cite{Frederix:2008hu,Frederix:2010cj},
which makes use of the dipole formalism as developed in \cite{Catani:1996vz}. For integration over
the phase space we used MadEvent \cite{Maltoni:2002qb}. The routines for generating histograms and distributions
originate from the MadAnalysis package (see http://madgraph.hep.uiuc.edu).

The generation of the various pieces and their 
combination for the phase space integration has been done in a fully automated way.

\subsection{Calculation of the virtual corrections}

The virtual corrections have been calculated with the automated one-loop 
amplitude generator {\sc GoSam}\,\cite{Cullen:2011ac}.
The program package {\sc GoSam} starts from an input card 
edited by the user and generates the diagrams 
and the corresponding expressions for the loop amplitudes in an automated way,
using {\tt QGRAF}\,\cite{Nogueira:1991ex}, {\tt FORM}~\cite{Vermaseren:2000nd,Kuipers:2012rf}
supplemented with {\tt Spinney}\,\cite{Cullen:2010jv} for the spinor algebra, 
and {\tt haggies}\,\cite{Reiter:2009ts} for the automated code generation.
It combines unitarity-inspired integrand reduction 
techniques~\cite{Ossola:2006us,Ellis:2007br,Mastrolia:2008jb,Mastrolia:2010nb,Heinrich:2010ax} 
with traditional tensor reduction methods\,\cite{Binoth:2008uq,Cullen:2011kv}.
The rational part can be calculated algebraically within {\sc GoSam} in an automated way.

We generated two versions of the virtual contributions, with and without top loops and 
found the effects of virtual top quarks negligible. Therefore the results presented here
have been obtained without the top contributions. Furthermore, we neglect one loop
contributions with two initial state gluons, which are formally of higher order, as 
 there is no corresponding tree level amplitude. Their contribution may 
potentially be enhanced by the large gluon luminosity~\cite{Binoth:1999qq,Bern:2002jx}. 
Their inclusion is however beyond the scope of this paper.

The remaining contributions can be reduced to the virtual corrections for the process
\begin{equation}
 q \bar{q} \to \gamma \, \gamma \, g\, .
 \label{virtual}
\end{equation}
The complete set of virtual corrections can be obtained from eq.\,(\ref{virtual}) by crossing
of the momenta and/or changing overall factors for different electromagnetic charges.
In total there are $130$ diagrams to be calculated up to pentagons.
The virtual amplitude has been checked for gauge invariance by 
adding a momentum dependent part to the 
photon polarisation vectors.


\subsection{Calculation of real radiation contributions}
As explained in detail in Section~\ref{sec:frag} above, 
the QCD corrections for processes involving photons contain 
infrared singularities related to the collinear emission of the 
photon off a final-state QCD parton. These singularities are 
compensated by the mass factorization 
terms of the photon fragmentation functions. 

From the computational point of view this implies that the real emission matrix element
contains collinear singularities which need to be regularized by some kind of subtraction terms.
The corresponding integrated subtraction terms make this singularity apparent as they develop
an explicit pole term  when integrated over the unresolved one-particle phase space in
dimensional regularisation. This pole is then absorbed into the fragmentation
functions.
To regulate these singularities we again make use of the dipole formalism 
as developed in \cite{Dittmaier:1999mb}
and implemented in the QED extension of MadDipole \cite{Gehrmann:2010ry}. This extension also
offers the framework for a straightforward implementation of fragmentation functions.
We refer to \cite{Gehrmann:2010ry} for further details. 

A collinear singularity between a photon and a quark can be regulated by a single dipole.
Note that in principle the role of emitter and unresolved particle is interchanged. In a true
QED calculation the quark would be the emitter and the photon the unresolved particle.
However in this calculation the photon is a tagged particle whereas the additional jet
can be unresolved. Nevertheless the dipole formalism for QED can be used as the subtraction
terms are symmetric under the exchange of emitter and unresolved particle if both are massless.
A small modification has been made in such a way that also photons are allowed as spectators.
This ensures that one can always use final-final dipole configurations, thereby reducing
the complexity of the calculation. 

Upon phase space integration, the pointwise cancellation of the  infrared poles from the virtual amplitude 
with those from the real radiation part has been  checked.
For both QCD and QED subtraction terms we have checked the independence on the $\alpha$-parameter
\cite{Nagy:1998bb}, which restricts the phase space segments on which  dipole subtraction 
is performed to the vicinity of the infrared singularities. 
 To perform this check we had to extend the usage of the $\alpha$-parameter
to the QED subtraction terms for non-collinear safe observables. In the purely massless case
and for final-final configurations this extension is straightforward.  

NLO corrections to $\gamma\gamma$+jet final states for the Frixione isolation 
criterion are free of collinear quark-photon contributions and do not 
depend on the photon fragmentation function. For this specific isolation 
criterion, NLO results for $\gamma\gamma$+jet production have been derived 
in~\cite{DelDuca:2003uz}. We fully reproduce these results, thereby 
obtaining a strong check on the correctness of our implementation of the 
virtual and real matrix elements, as well as on the non-QED-type 
subtraction terms. 

\section{Numerical results}
\label{sec:results}

In this section we present phenomenological results 
for proton-proton collisions at  $\sqrt{s}=8$\,TeV.
The results are divided into two categories: the one-jet inclusive case, $pp\to \gamma\gamma$+jet+$X$, 
and the one-jet exclusive case  $pp\to \gamma\gamma$+jet. The inclusive sample is defined 
by requiring that at least one jet in the event passes the selection cuts introduced below, while 
the exclusive sample admits only events containing exactly one jet within the selection cuts.

\subsection{Input parameters and kinematic cuts}

For the  jet clustering we used an anti-$k_T$ algorithm~\cite{Cacciari:2008gp} with a cone size
of $R_j=0.4$ provided by 
the {\tt FastJet} package \cite{Cacciari:2011ma,Cacciari:2005hq}.
We used an NLO 
parton distribution set from  NNPDF2.3 \cite{Ball:2012cx}, where the values for $\alpha_s$ at leading order 
and next-to-leading order are given by
$$
\alpha_{s}(M_Z) = 0.119\;, 
$$
and the running is calculated at one loop for the LO results and at two loops for the NLO
 results. 
For the photon fragmentation functions, we take set II of the 
parametrisations of Ref.~\cite{Bourhis:1997yu}.

The following kinematic cuts have been applied:
$p_T^{\rm{jet}}>40$\,GeV, $p_T^{\gamma}>20$, $|\eta^{\gamma},\eta^j| \leq 2.5$, $R_{\gamma ,j} > 0.4$, $R_{\gamma, \gamma} >0.8$
and 100\,GeV $\leq m_{\gamma\gamma} \leq $ 140\,GeV. The intention of the latter cut is to 
focus on a region around the Higgs resonance.

For the photon isolation, we compare the Frixione isolation criterion with the fixed cone criterion
for several values of the photon energy fraction $\epsilon_c$ in the cone.
For the Frixione isolation criterion (see eq.~(\ref{eq:frix})), 
our default values are $R=0.4, n=1$ and $\epsilon=0.5$. 
For the cone-based isolation, the default cone size is $R=0.4$, 
while several different values of the hadronic energy fraction $z_c$ inside the cone will be used, 
where $$z_c=\frac{|\vec{p}_{T,{\rm cone}}^{\rm{\,had}}|}{|\vec{p}_T^{\,\gamma}+\vec{p}_{T,{\rm cone}}^{\rm{\,had}}|}\;,$$ 
such that in the collinear limit,  $z_c$ is related to $\epsilon_c$ in   eq.~(\ref{eq:cone-iso}) by $z_c=\frac{\epsilon_c}{1+\epsilon_c}$.

\subsection{Scale dependence and sensitivity to the isolation parameters}

The truncation of the perturbative expansion of a collider observable leads to a dependence 
on scale parameters that were introduced in the renormalization and mass factorization. 
The residual dependence on these parameters is often used to quantify the 
uncertainty on the calculation from missing higher order terms in the 
perturbative series. 
Apart from the dependence on renormalization and initial state factorisation scales $\mu_r,\mu_f$, 
the cross section for the production of prompt photons also depends on the fragmentation scale $\mu_F$, 
as explained in section \ref{sec:frag} above.
To study the scale dependence of our NLO results, 
we set $\mu_r=\mu_f=\mu_F$ and  choose $\mu_0^2=\frac{1}{4}\,(m_{\gamma\gamma}^2+\sum_j p_{T,j}^2)$ for our central scale.  
The scales are then varied by $\mu=x\,\mu_0$ with $0.5\leq x \leq 2$. 

In Figure~\ref{fig:scalevarexcl}, we display the scale dependence of the exclusive $\gamma\gamma$+jet 
cross section. For a cone-based isolation criterion, we 
observe a clear reduction of the scale dependence at next-to-leading order, while this reduction is less 
pronounced for the Frixione isolation criterion. This qualitative difference can be attributed to 
the occurrence of almost collinear quark-photon configurations at NLO. The typical scale of these 
configurations is the invariant mass of the quark-photon system, which can be 
substantially lower than $\mu_0$. In the case of a cone-based isolation, these contributions combine with 
the fragmentation contribution, which compensates their scale dependence. For the Frixione  isolation 
criterion, this compensation does not occur, thereby resulting in a larger scale-dependence.

In the inclusive case, Fig.~\ref{fig:scalevarincl}, no reduction of the scale uncertainty at NLO is visible.
The reason for this is the fact that the cross section in this case is dominated by the $\gamma\gamma +$2\,jets
real radiation part, which shows a leading order scale dependence. A similar behaviour has been observed for 
example in  $ZZ+$jet production with and without second jet veto~\cite{Binoth:2009wk}.
The largeness of the NLO corrections has already has been observed for 
Frixione isolation in Ref.~\cite{DelDuca:2003uz}. 
The reasons are the appearance of new partonic 
subprocesses at NLO and the enlarged final state phase space at this order. 

\FIGURE{
\parbox{19.cm}{
\epsfig{file=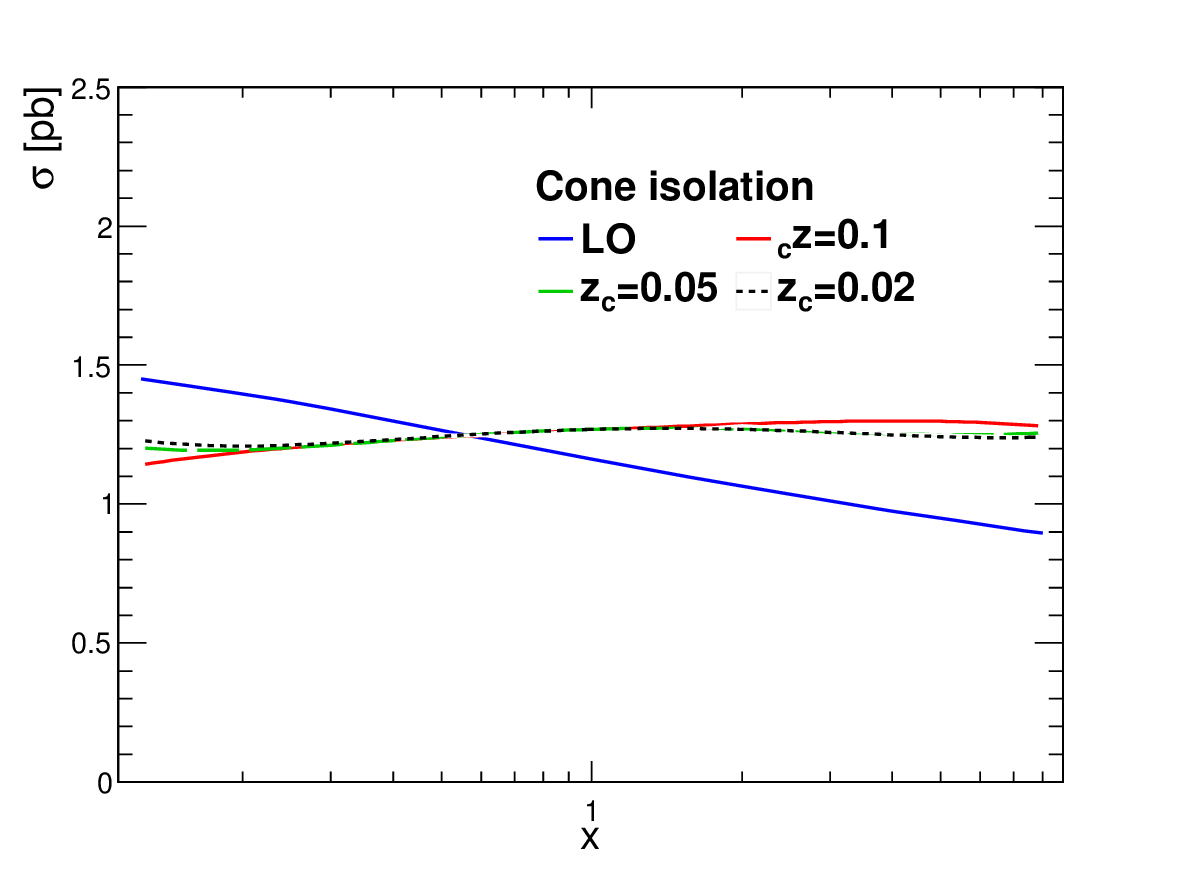,width=9.cm}
\epsfig{file=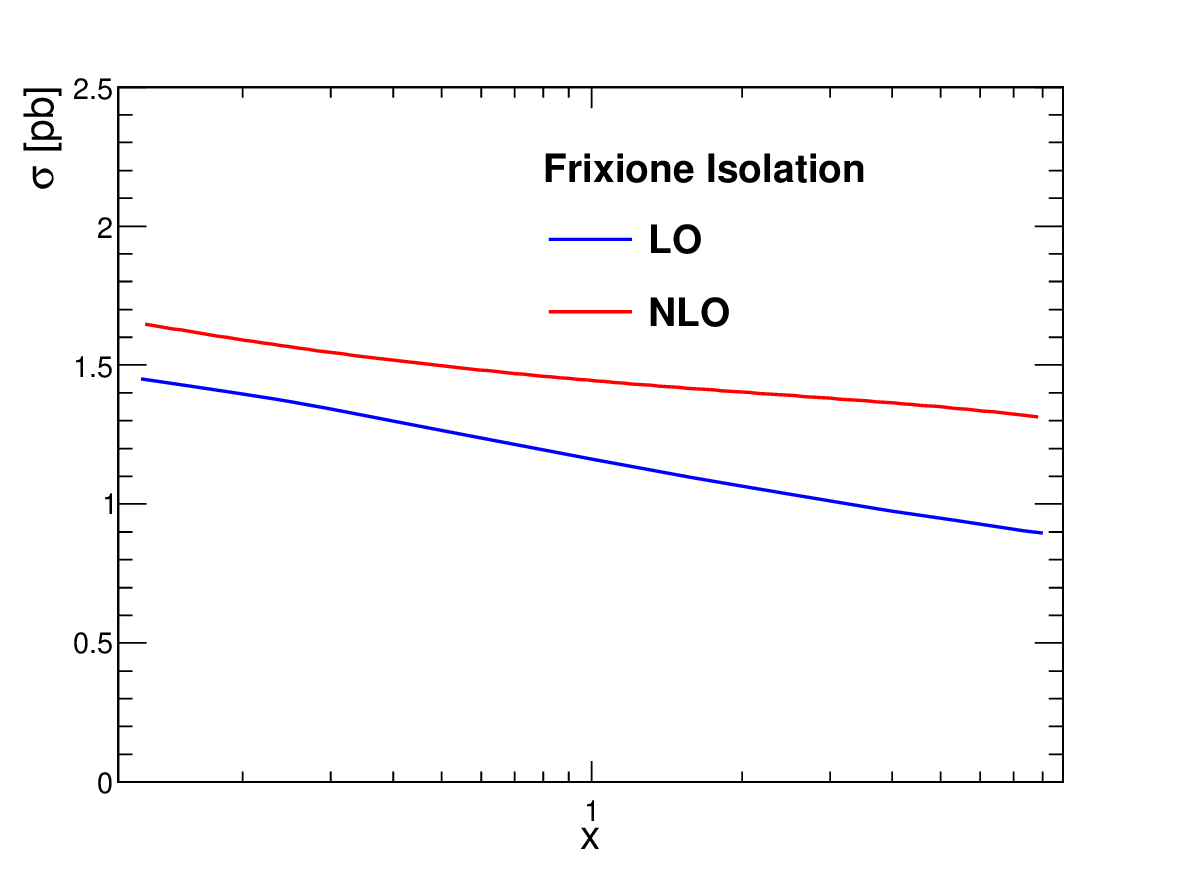,width=9.cm}
\caption{Behaviour of the exclusive $\gamma\gamma$+jet  cross sections with different isolation
prescriptions under scale variations, 
$\mu=x\,\mu_0$,  $0.5\leq x \leq 2$, $\mu_0^2=\frac{1}{4}\,(m_{\gamma\gamma}^2+\sum_j p_{T,j}^2)$.
 \label{fig:scalevarexcl} }
}
}

\FIGURE{
\parbox{19.cm}{
\epsfig{file=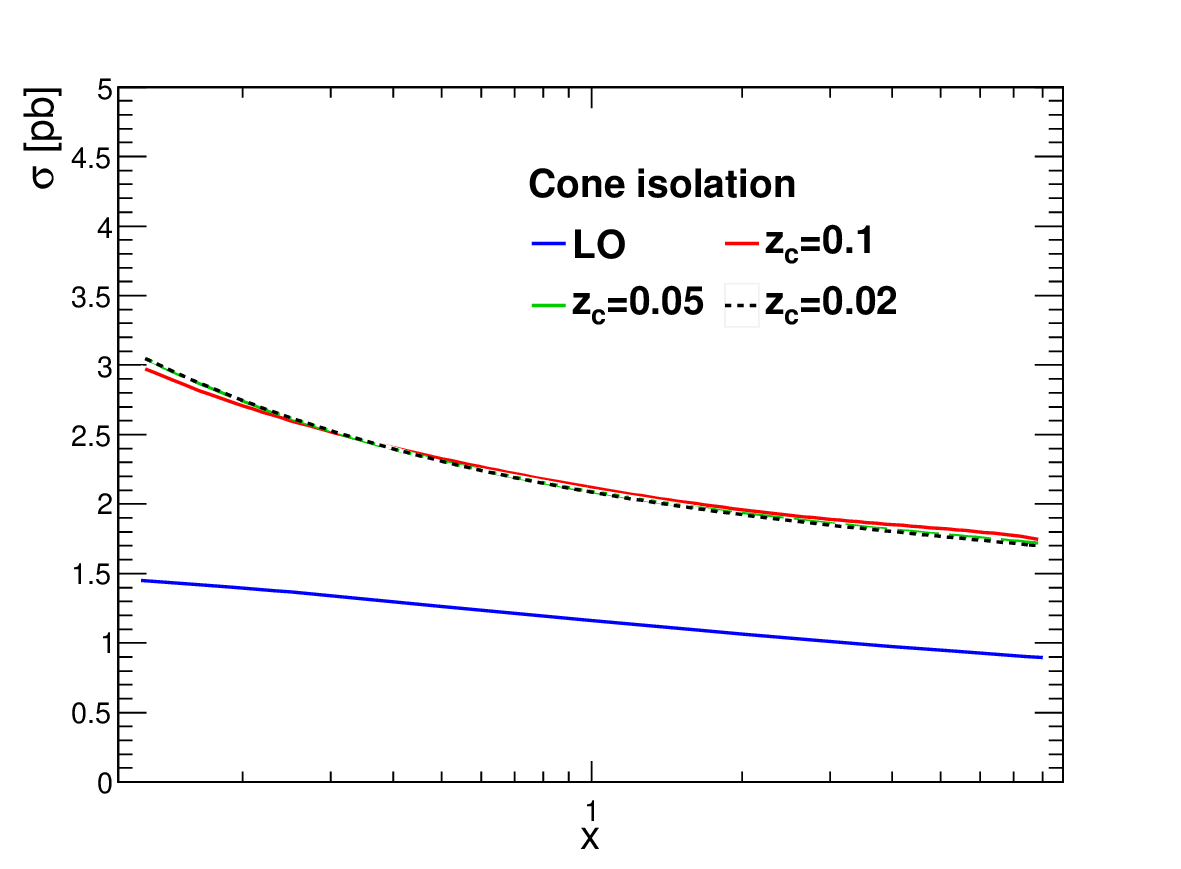,width=9.cm}
\epsfig{file=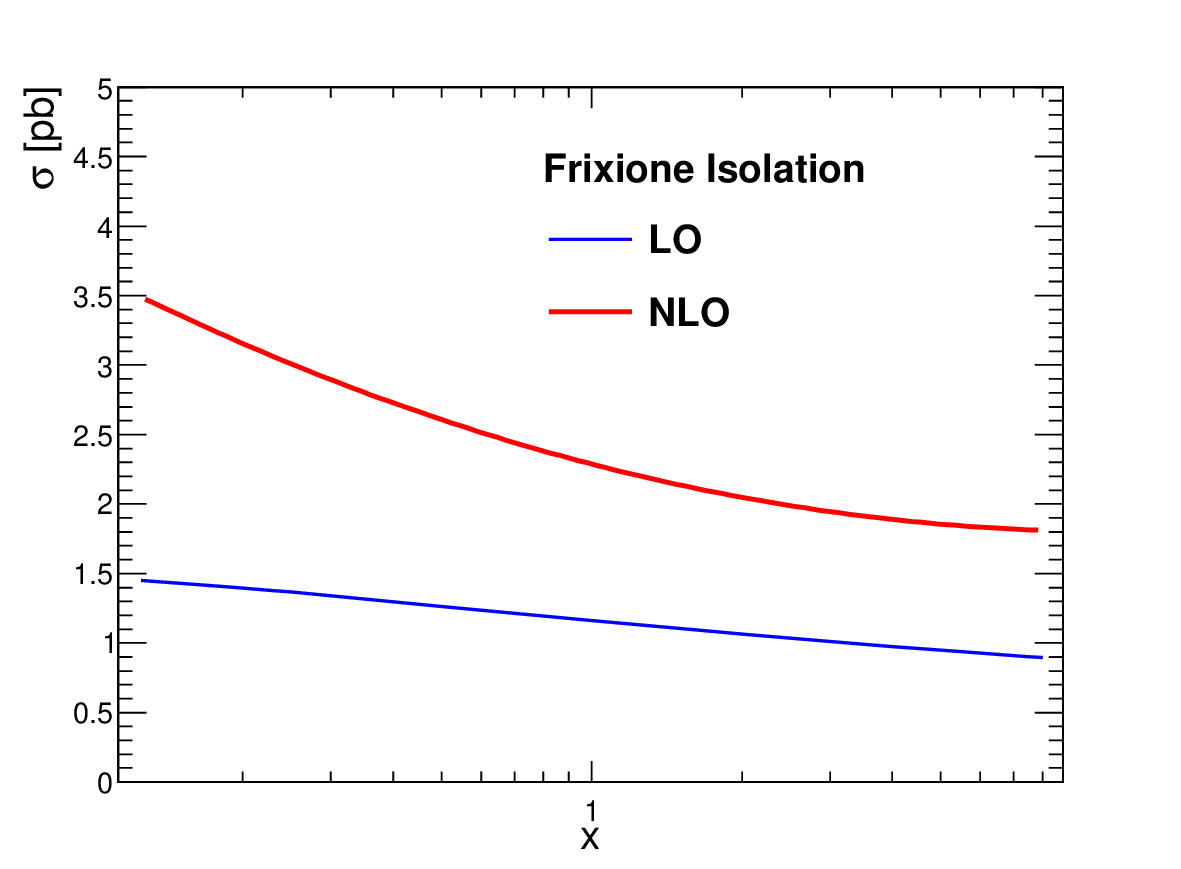,width=9.cm}
\caption{Behaviour of the inclusive $\gamma\gamma$+jet+X cross sections with different isolation
prescriptions under scale variations, 
$\mu=x\,\mu_0$,  $0.5\leq x \leq 2$, $\mu_0^2=\frac{1}{4}\,(m_{\gamma\gamma}^2+\sum_j p_{T,j}^2)$.
 \label{fig:scalevarincl} }}
}

Due to photon isolation, the cross section for  $pp\to \gamma\gamma$+jet+X is not strictly an inclusive quantity.
The integration over the final state collinear variable $z$ appearing in the fragmentation functions 
$D_q^\gamma(z,\mu_F) $ is bounded from below by $1-z_c$.

Further, the presence of three different scales $\mu_r,\mu_f,\mu_F$ partially leads to a scale 
dependence which is different from what is known from pure QCD.

Figures \ref{fig:compz} and \ref{fig:compFrix} show how the scale variation bands vary as a function of the 
isolation parameters, for both the single-jet inclusive and the exclusive case. 
One observes that for both isolation criteria the inclusive case is dominated by the large scale dependence of the $\gamma\gamma+2$\,jets 
part of the real radiation which has an uncompensated  leading order scale dependence.

 \FIGURE{
\begin{minipage}{10cm}
\includegraphics[width=9.cm]{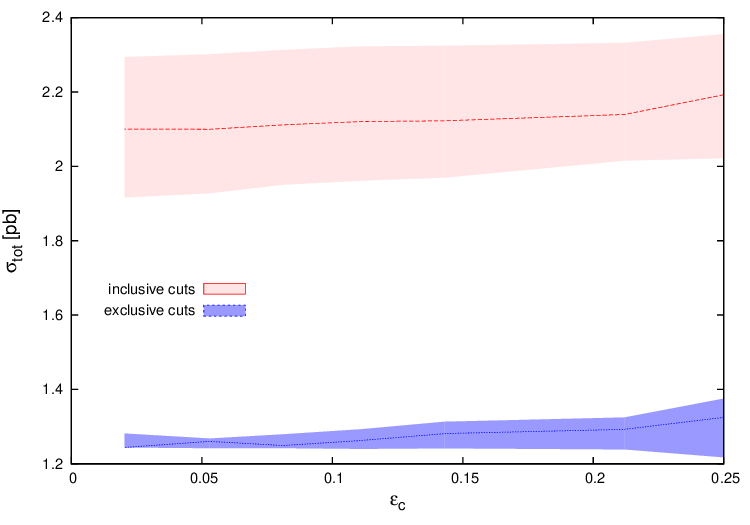} 
 \caption{Dependence of the cross sections on the cone isolation parameter $\epsilon_c$. 
 The bands correspond to scale variations $0.5\leq x \leq 2$, with $\mu=x\,\mu_0$.
 \label{fig:compz} }
\end{minipage}
}

 \FIGURE{
\begin{minipage}{10cm}
\includegraphics[width=9.cm]{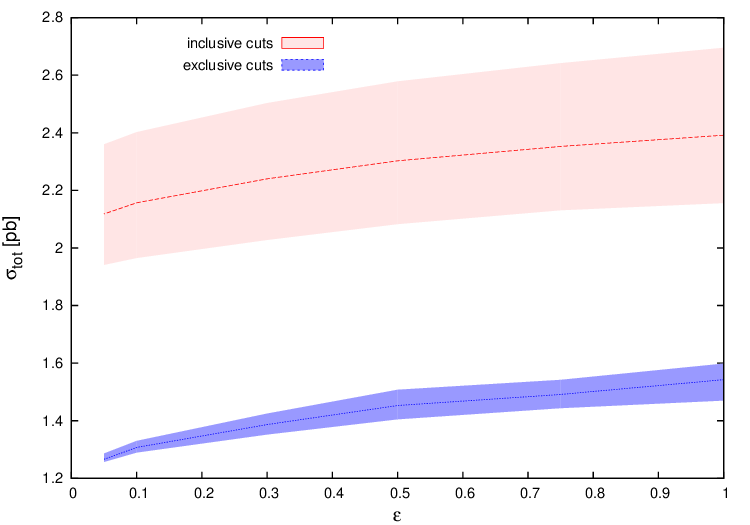} 
 \caption{Dependence of the cross sections on the Frixione isolation parameter $\eps$. \label{fig:compFrix} }
\end{minipage}
}

Comparing Fig.~\ref{fig:compz} and Fig.~\ref{fig:compFrix}, 
it can be seen that the qualitative dependence on the hadronic energy
threshold parameters 
$\epsilon_c$ (in fixed cone isolation) and $\epsilon$ (in Frixione isolation) is substantially different. 
In particular, we observe that the
cross section remains almost constant over a large range of  $\epsilon_c$, 
while for Frixione isolation, the cross section increases mildly over the interval
$0.1<\epsilon<1$. 
This difference in the qualitative behaviour shows that the parameters can not be translated
into each other, since the threshold in the fixed cone isolation is rigid, 
while the threshold in the
Frixione isolation is dynamical, and weighted by the distance to the cone axis. 

The pattern of the scale variation band for $\epsilon_c$ in the case of exclusive cuts 
with cone isolation can be understood from the fact that 
there are several cancellations of scale dependent terms at work. 
The renormalisation scale dependence acts in opposite direction to the fragmentation scale dependence, 
i.e. the cross section decreases with increasing $\mu_r$, while it increases with increasing $\mu_F$.
The $\log(\mu_F)$ terms which are contained in the perturbative component of the fragmentation functions 
largely cancel the $\mu_F$ dependence of the direct real radiation part. 
Only for larger $\epsilon_c$ values non-perturbative and beyond-leading-logarithmic effects start to 
become important. 

With Frixione isolation, the $\log(\mu_F)$ dependence is absent, therefore the behaviour 
under scale variations in Fig.~\ref{fig:compFrix} 
is qualitatively different.

\subsection{Results for diphoton plus one jet production -- exclusive case}

In the exclusive case, only moderate NLO corrections are observed in the kinematical distributions 
related to $\gamma\gamma$+jet final states. Figure~\ref{fig:mggexcl}  displays the 
photon pair invariant mass and the jet transverse momentum distributions. We observe that the 
NLO corrections are in general limited in magnitude, and slightly larger for  
the Frixone isolation than for the cone-based isolation.  Inclusion of the 
NLO corrections amounts to an rescaling of the distributions that is constant in invariant mass, and 
slightly decreasing with jet transverse momentum. 

\FIGURE{
\parbox{16.cm}{
\epsfig{file=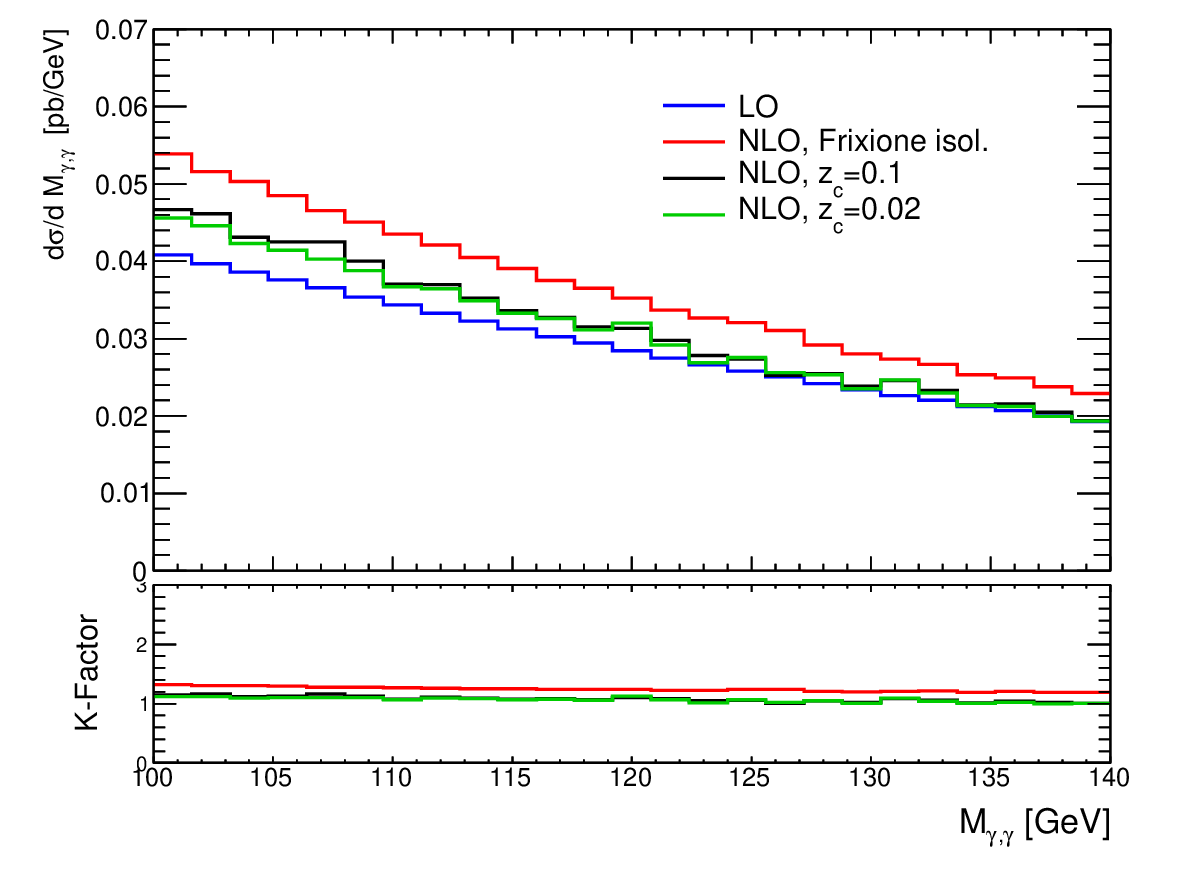,width=7.9cm}
\epsfig{file=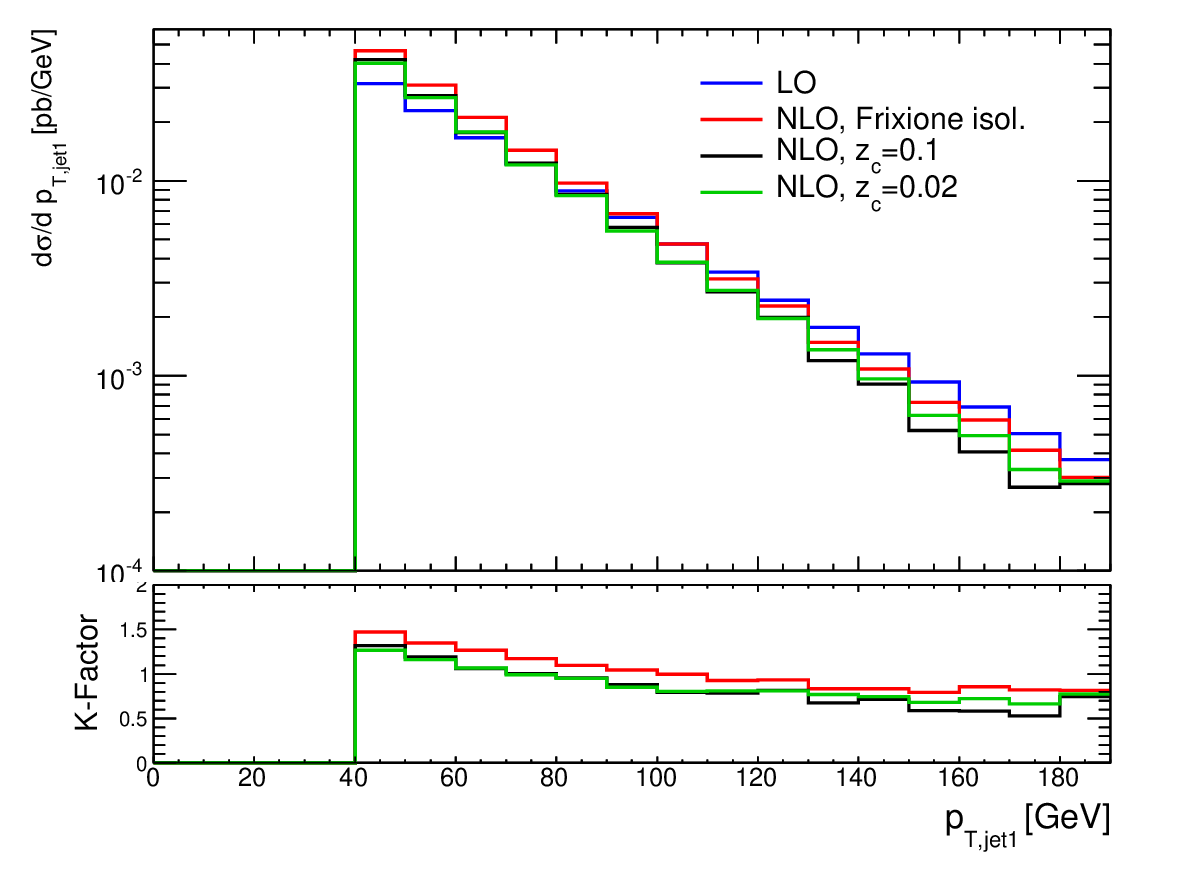,width=7.9cm}
\caption{(a) Photon invariant mass distribution, (b) transverse momentum distribution of the leading jet
for the diphoton plus one jet exclusive cross section.
 \label{fig:mggexcl}} 
}
}
In Figure \ref{fig:ptaexcl}, the transverse momentum distributions of the leading and the subleading photon are 
displayed. It can be seen that the leading-$p_T$ photon tends to become softer at NLO, 
which is natural since the additional QCD radiation carries away momentum.

\FIGURE{
\parbox{16.cm}{
\epsfig{file=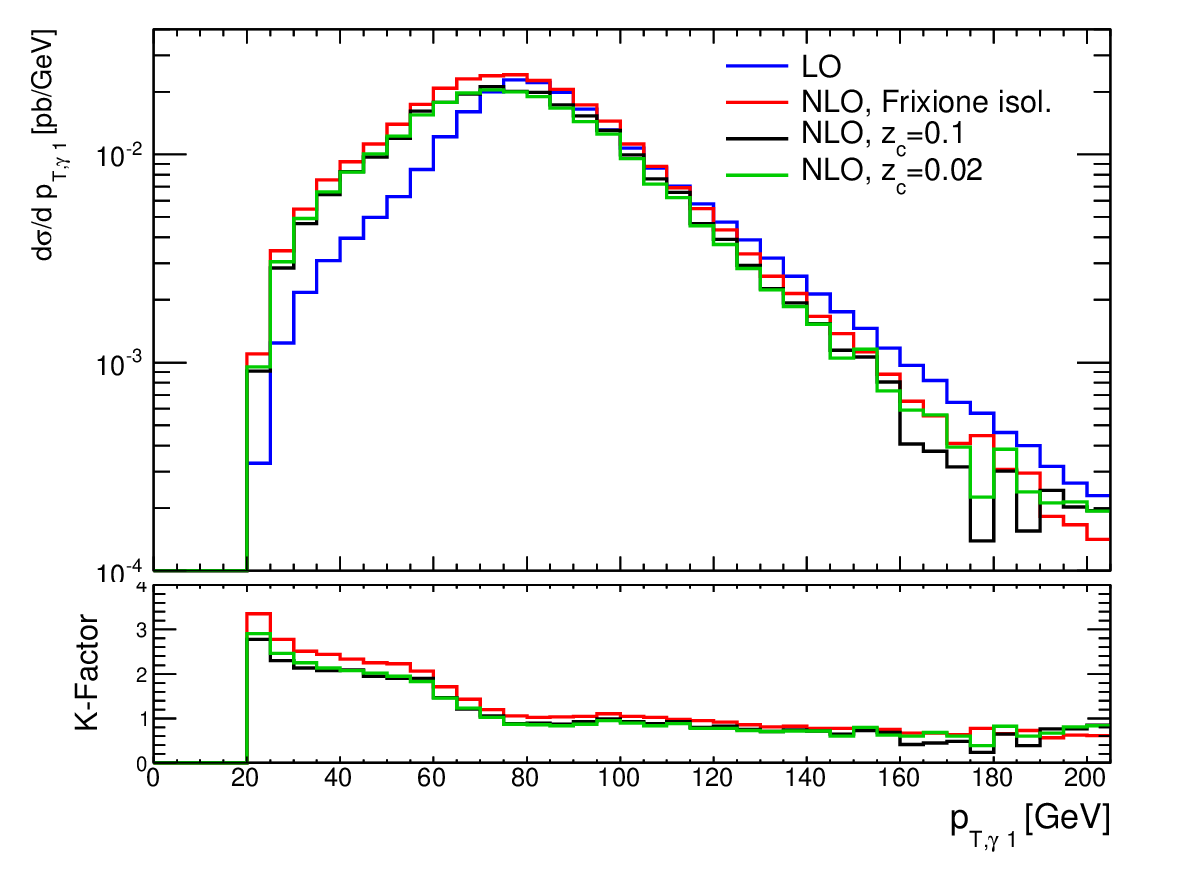,width=7.9cm}
\epsfig{file=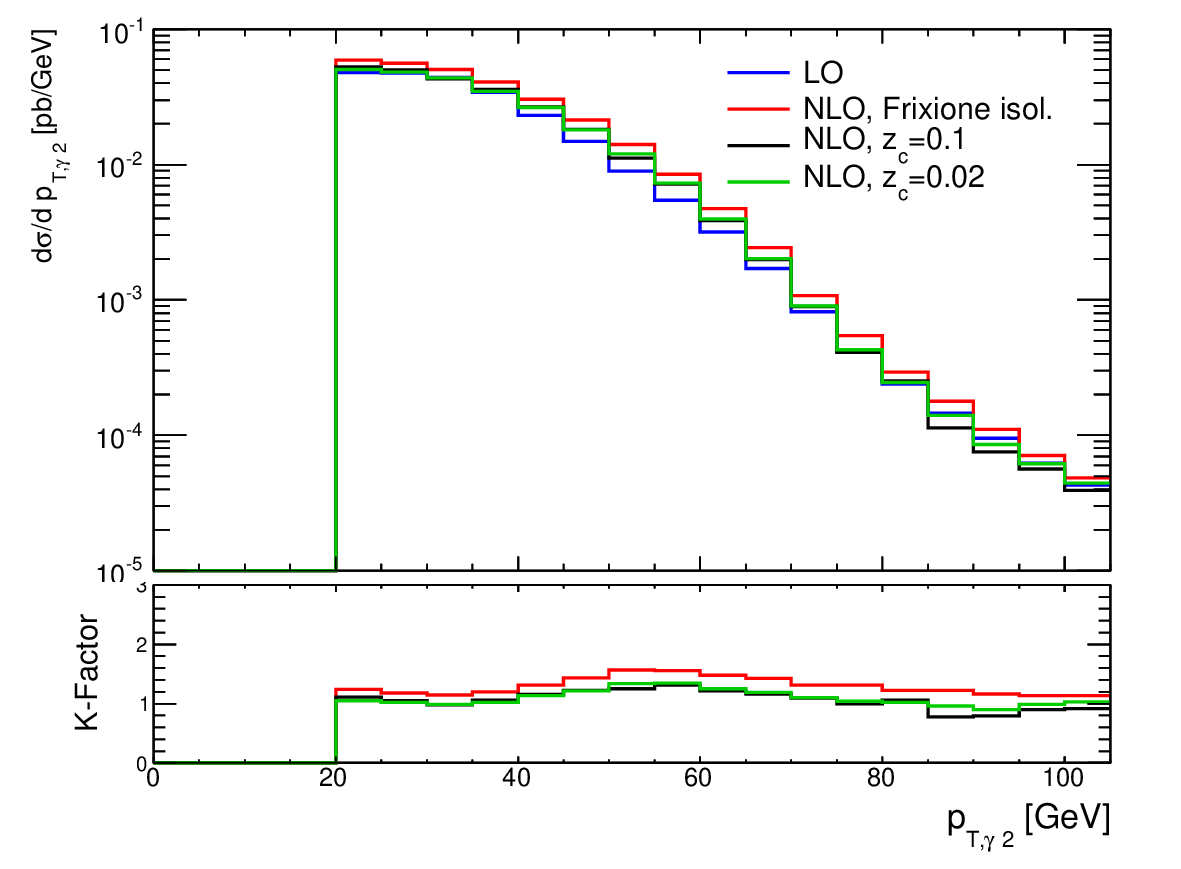,width=7.9cm}
\caption{Transverse momentum distributions of the leading (i.e. larger-$p_T$) and subleading photon
for the diphoton plus one jet exclusive cross section.
 \label{fig:ptaexcl}} 
}
}
Figure \ref{fig:Rjetgexcl} shows the differential distributions in the distance 
$R_{j\gamma} = \sqrt{(\eta^j-\eta^{\gamma})^2-(\phi^j-\phi^{\gamma})^2}$ between the
jet and the harder photon ($\gamma_1$) respectively the 
softer photon ($\gamma_2$), with a separation cut of $R_{j\gamma} \geq 0.4$.
At leading order, the preferred kinematical configuration of the jet is 
back-to-back with the harder photon, and near to the softer photon. These
kinematical correlations are weakened with the inclusion of 
NLO corrections, with an increase of events with smaller 
opening angle between the hard photon and the jet or larger opening 
angle between the soft photon and the jet. 
\FIGURE{
\parbox{16.cm}{
\epsfig{file=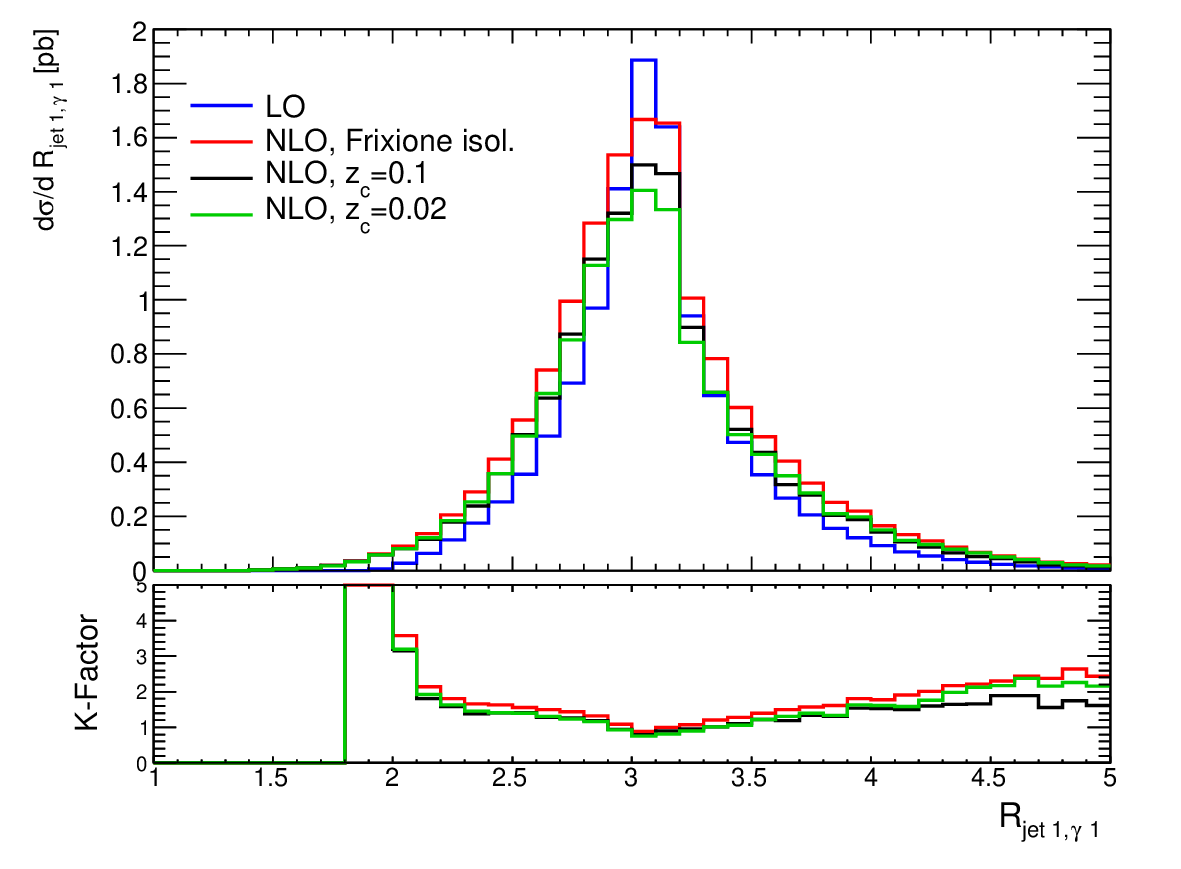,width=7.9cm}
\epsfig{file=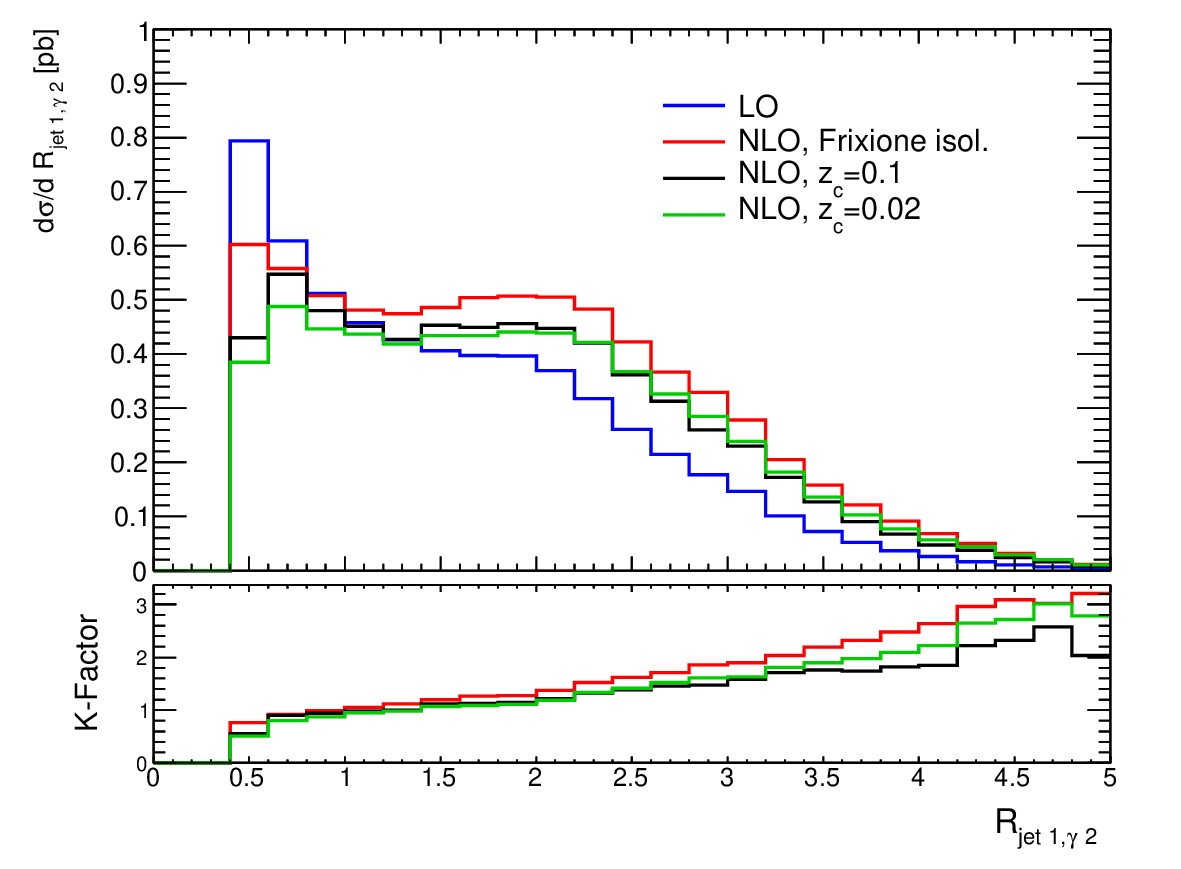,width=7.9cm}
\caption{$R$-separation between photon and jet in the $\eta-\phi$ plane 
for the diphoton plus one jet exclusive cross section.
$R_{jet_1,\gamma_1}$ denotes the $R$-separation between the jet and the harder photon,
while $R_{jet_1,\gamma_2}$ is the $R$-separation between the jet and the softer photon.
 \label{fig:Rjetgexcl}} 
}
}

\clearpage

\subsection{Results for diphoton plus one jet production -- inclusive case}

As already observed for the total cross section, NLO corrections are 
substantially larger for the inclusive cross section $\gamma\gamma+$jet+$X$ as 
compared to the exclusive case. In the inclusive cross section, the substantial 
contribution from  $\gamma\gamma+2$~jet final states results in larger 
corrections, and induces substantial modifications to some of the kinematical distributions.

Figure~\ref{fig:mggincl} displays the inclusive distributions in photon pair invariant mass 
and leading jet transverse momentum. The magnitude of the corrections is larger than in 
the exclusive case, they remain constant for the invariant mass distribution and rise 
with the jet transverse momentum (as opposed to the decrease with jet transverse momentum 
in the exclusive case, Figure~\ref{fig:mggexcl}). Again, the corrections for the Frixione isolation 
criterion are slightly larger than for the fixed-cone isolation. 
\FIGURE{
\parbox{16.cm}{
\epsfig{file=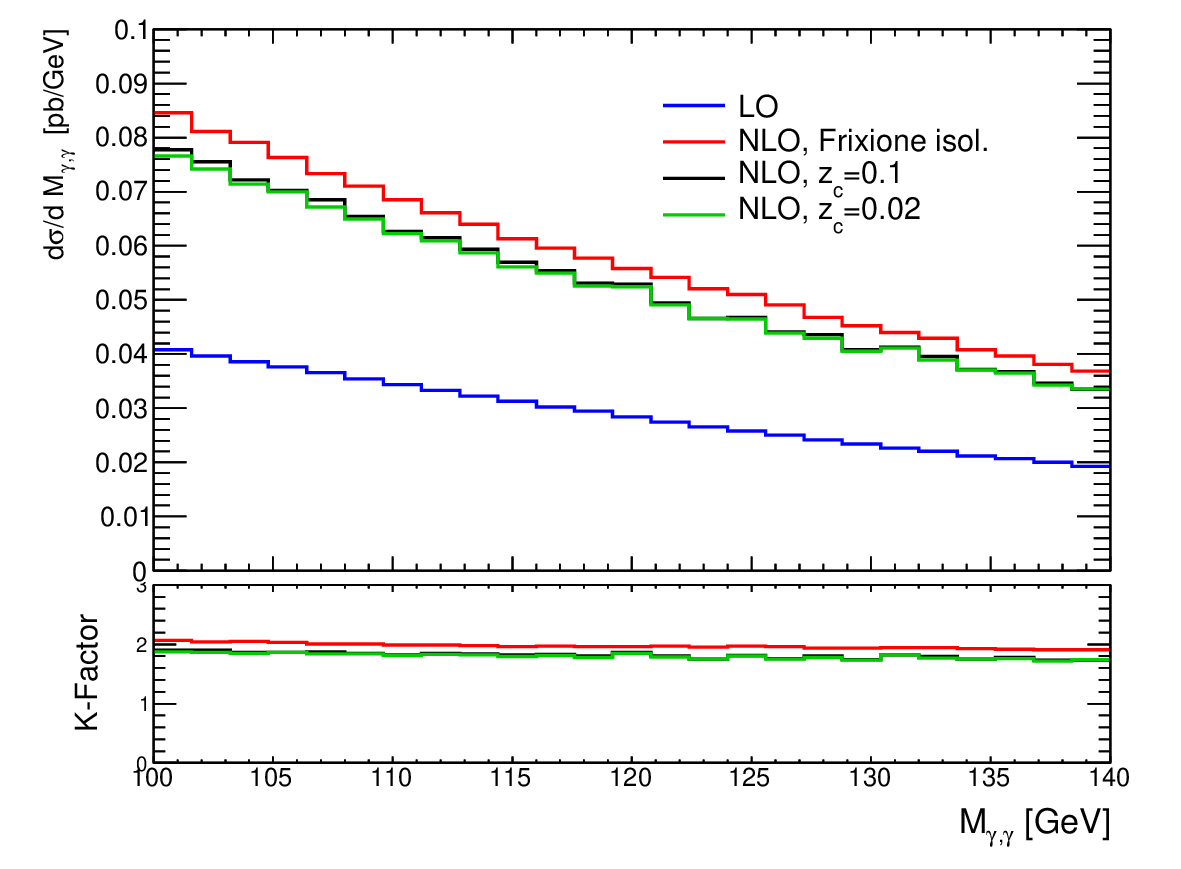,width=7.9cm}
\epsfig{file=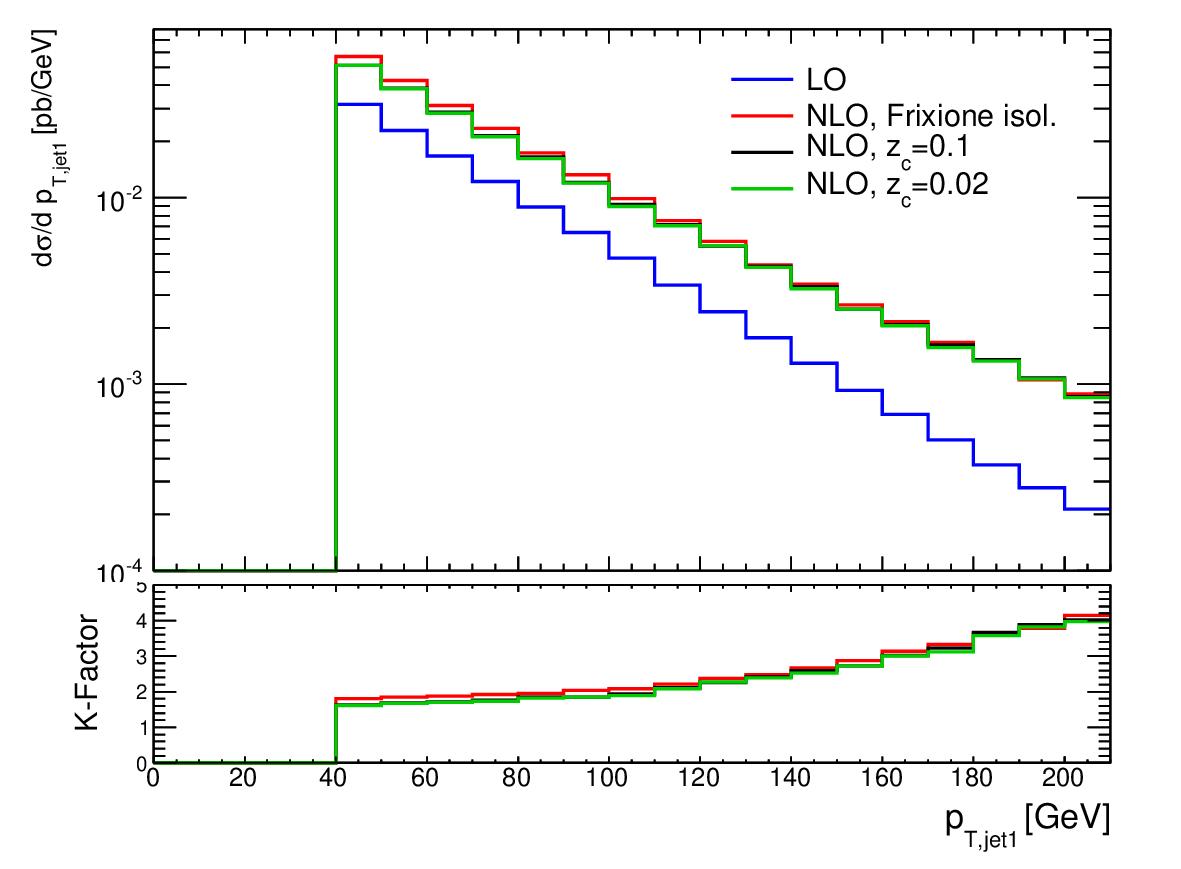,width=7.9cm}
}
\caption{(a) Photon invariant mass distribution, (b) transverse momentum distribution of the leading jet
for the diphoton plus one jet inclusive cross section.\label{fig:mggincl}}
}

The photon transverse momentum distributions, Figure~\ref{fig:ptaincl}, display a similar 
behaviour as in the exclusive case, with the main effect from NLO corrections appearing in 
a softening of the leading photon distribution. 
The effect of the extra jet in the inclusive distribution is particularly pronounced in the 
$R_{j\gamma}$ distributions. 
\FIGURE{
\parbox{16.cm}{
\epsfig{file=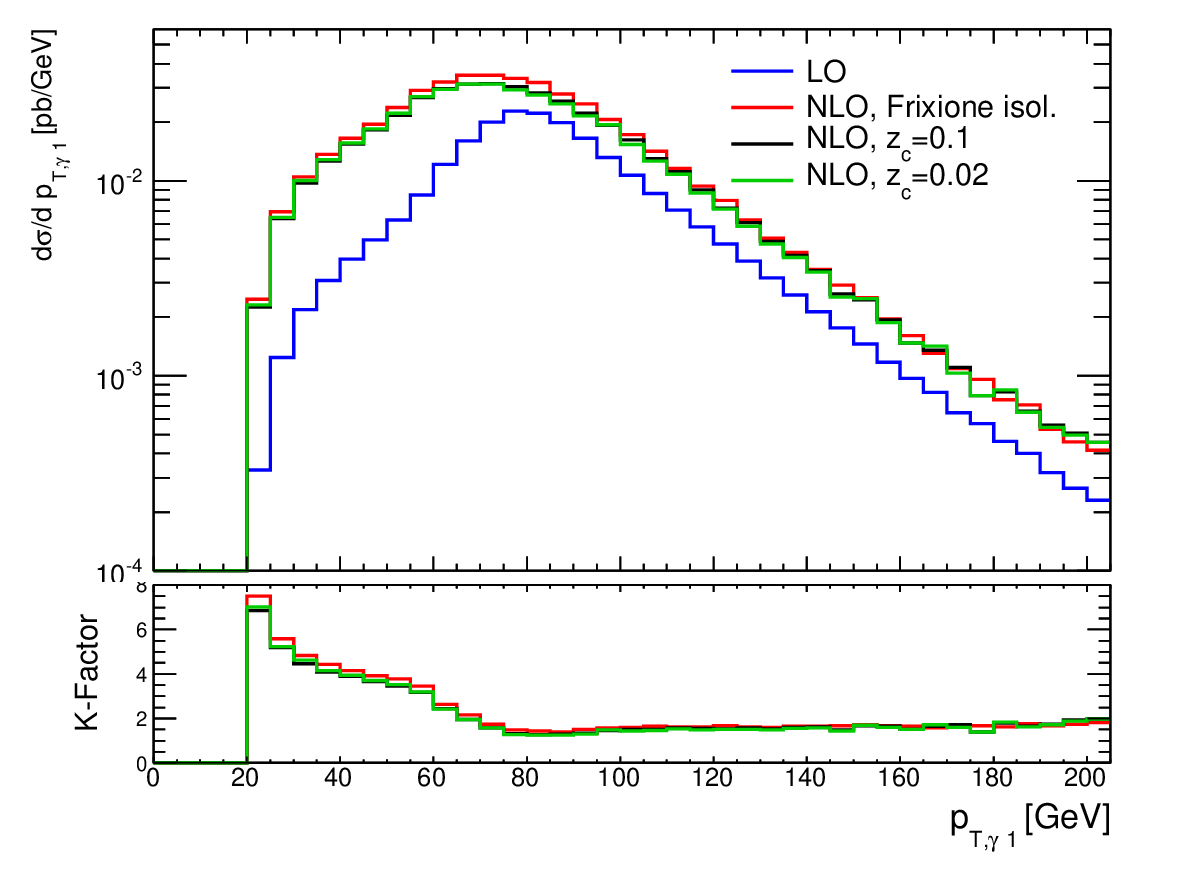,width=7.9cm}
\epsfig{file=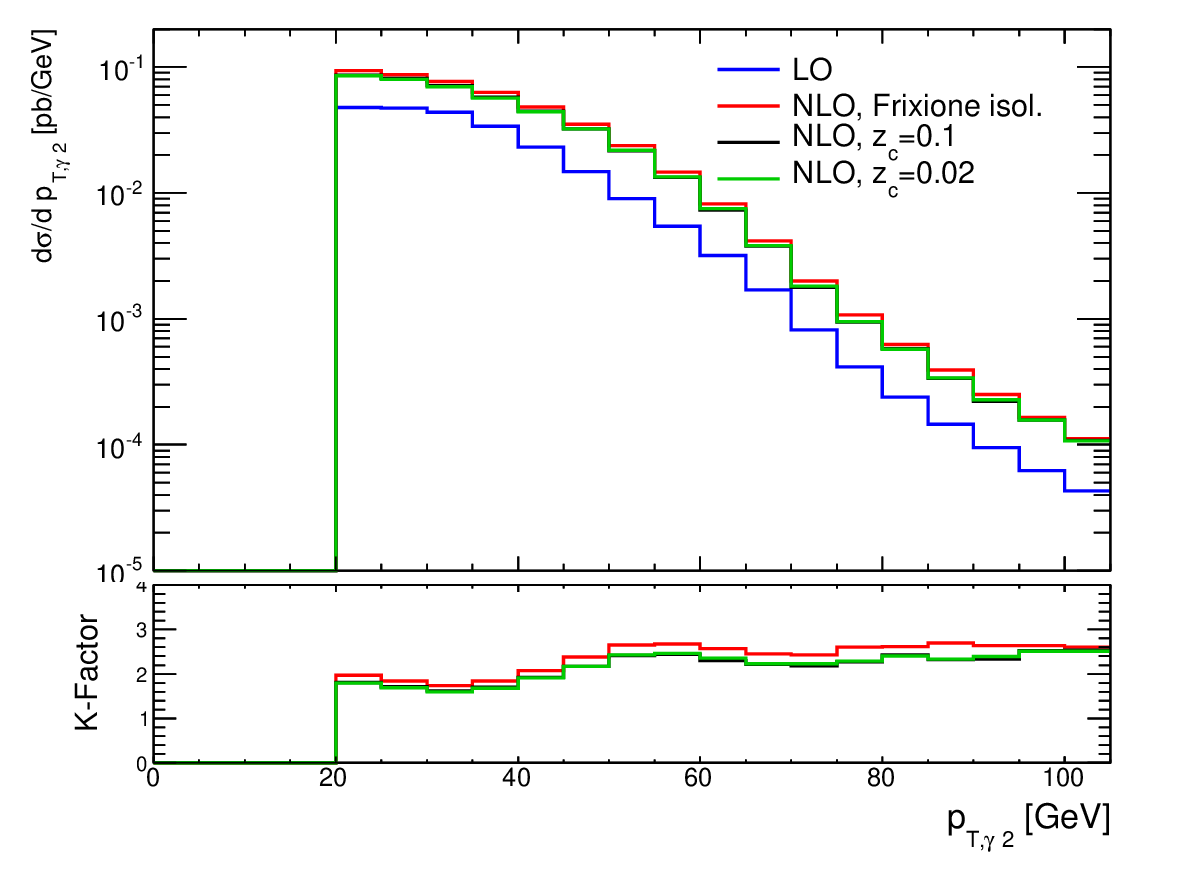,width=7.9cm}
}
\caption{Transverse momentum distributions of the leading (i.e. larger-$p_T$) and subleading photon
for the diphoton plus one jet inclusive cross section.
 \label{fig:ptaincl}} 
 }

\FIGURE{
\parbox{16.cm}{
\epsfig{file=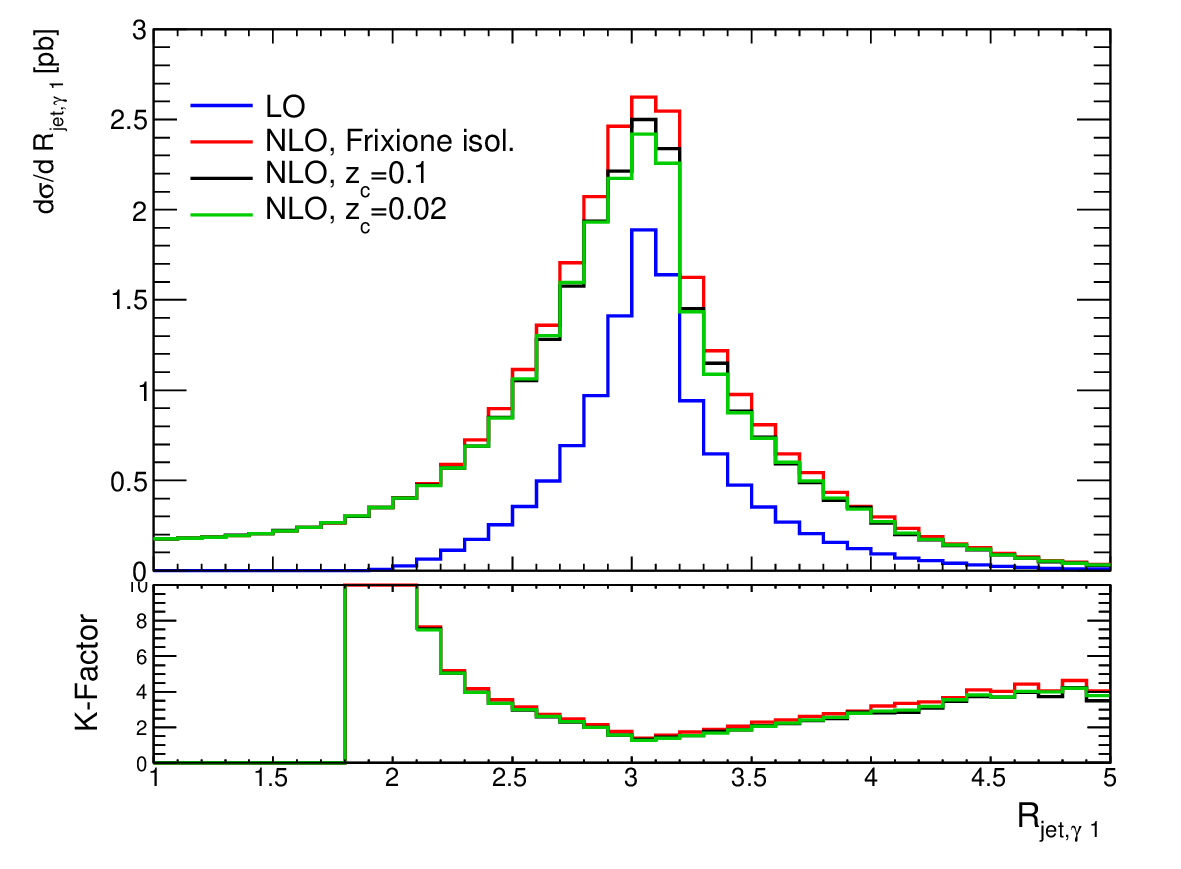,width=7.9cm}
\epsfig{file=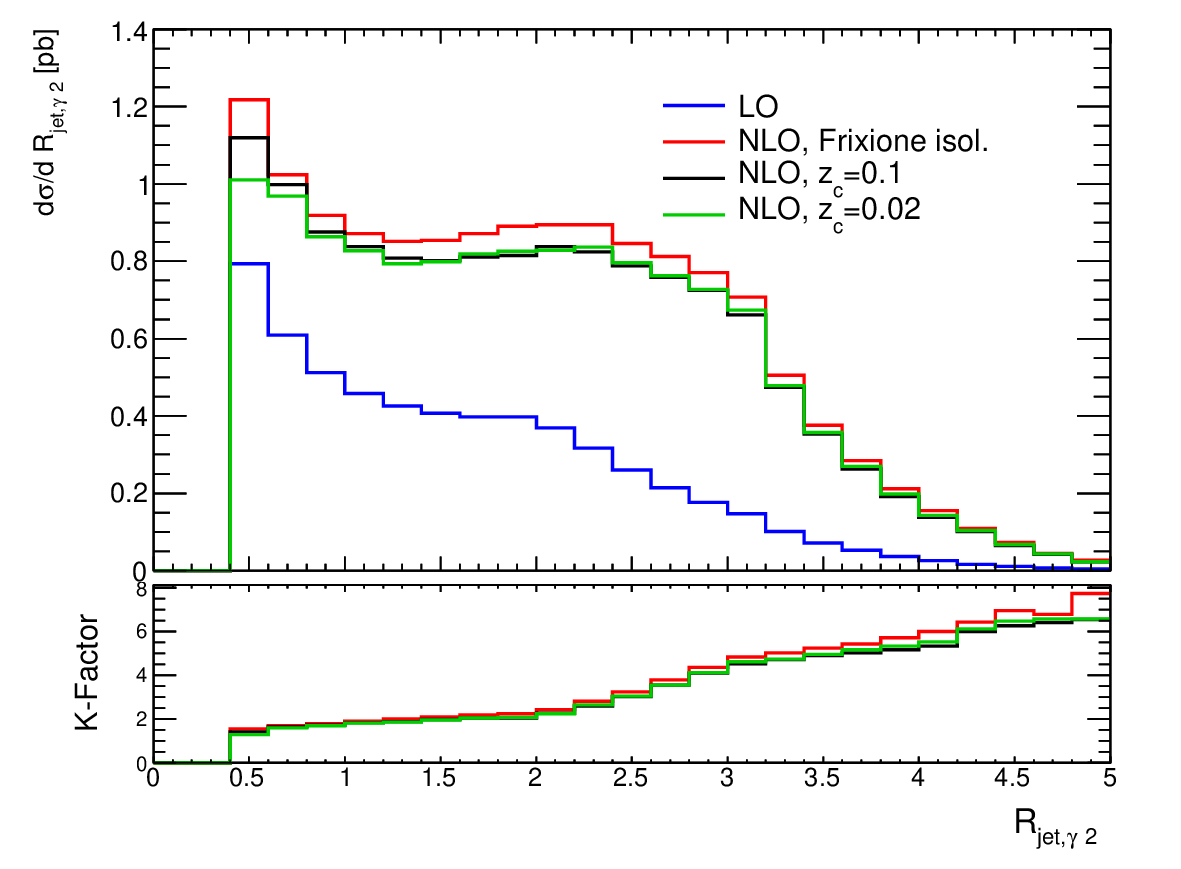,width=7.9cm}
}
\caption{$R$-separation between photon and jet in the $\eta-\phi$ plane 
for the diphoton plus one jet inclusive cross section.
$R_{jet_1,\gamma_1}$ denotes the $R$-separation between the jet and the harder photon,
while $R_{jet_1,\gamma_2}$ is the $R$-separation between the jet and the softer photon.
 \label{fig:Rjetgincl}} 
}

Comparing exclusive (Fig.~\ref{fig:Rjetgexcl}) and inclusive (Fig.~\ref{fig:Rjetgincl})
cases, one can see very clear differences. 
For example, in the first bins of the $R_{j\gamma_1}$ distribution 
(separation between leading jet and harder photon), the inclusive case shows a shoulder
due to the contributions from the second jet, which is vetoed in the exclusive case.
Further, in the first bins of the $R_{j\gamma_2}$ distribution 
(separation between leading jet and softer photon), the K-factor is smaller than one in the exclusive case, 
while it is always larger than unity in the inclusive case. Note however that in the inclusive case, 
events where both jets fulfill the cuts are counted twice. Therefore it is somewhat misleading to 
directly compare K-factors between the inclusive and exclusive case.

\clearpage

\section{Conclusions and outlook}
\label{sec:conclusion}

We have calculated the NLO QCD corrections to the production of two isolated photons 
in association with a jet at hadron colliders. 
Results for both the one-jet inclusive case as well as the case where 
exactly one jet passing the jet selection criteria have been presented.
Our calculation also includes contributions from the fragmentation of a hadronic 
jet into a large-$p_T$ photon at order $\alpha^2\alpha_s^2$, 
and therefore allows to compare different photon isolation criteria. 
Comparing the Frixione isolation criterion to standard cone isolation 
for several values of the hadronic energy allowed in the cone, we observe that 
the scale dependence stabilizes at NLO for the exclusive cross section in the 
standard cone isolation case, while with Frixione isolation
the stabilisation is much less pronounced. 
This behaviour can be attributed to the fact that 
there are cancellations of logarithms stemming from the 
factorisation of collinear quark-photon splittings between the direct and the fragmentation 
contributions. For very strict isolation parameters, the results for standard cone 
and Frixione isolation are similar with regards to the scale dependence. 
The K-factors are in general larger with Frixione isolation than with cone isolation.

In contrast to the exclusive cross section, the  one-jet inclusive cross section 
does not show a stabilisation of the scale dependence, independent of the choice of the isolation criterion.
This can be understood from the fact that the cross section in this case is dominated by the 
two-jet contribution of the 
NLO real radiation part, which shows a leading order scale dependence.

The code which is underlying the calculation presented here 
has been made publicly available at {\tt http://gosam.hepforge.org/diphoton}. 
This will allow further dedicated studies of different isolation 
prescriptions and kinematic situations, and eventually the combination 
of the NLO calculation with a parton shower.

\section*{Acknowledgements}
We would like to thank the {\sc GoSam} collaboration for helpful discussions.
N.G. and G.H. would like to thank the University of Zurich for kind hospitality, 
where parts of the project were carried out.
This research is supported in part by
the Swiss National Science Foundation (SNF) under contract
200020-138206 and  by the European Commission through the 
``LHCPhenoNet" Initial Training Network PITN-GA-2010-264564.

\section*{Note added}
After publication of this paper, diphoton production in association with a jet, 
including the fragmentation component, 
was re-calculated in Ref.~\cite{Campbell:2014yka}. After communication with the authors of 
Ref.~\cite{Campbell:2014yka} it turned out that the original version of our code did not 
call the correct photon fragmentation functions. Even though the effect is minor, 
except in Figure \ref{fig:compz}, we have replaced all figures with new ones 
based on the corrected version of our code.
We thank the authors of Ref.~\cite{Campbell:2014yka} for pointing us towards this problem.


\begin{thebibliography}{10}

\bibitem{:2012gk}
{\bf ATLAS} Collaboration, G.~Aad et~al., {\it {Observation of a new particle
  in the search for the Standard Model Higgs boson with the ATLAS detector at
  the LHC}},  {\em Phys.Lett.} {\bf B716} (2012) 1--29,
  [\href{http://xxx.lanl.gov/abs/1207.7214}{{\tt arXiv:1207.7214}}].

\bibitem{:2012gu}
{\bf CMS} Collaboration, S.~Chatrchyan et~al., {\it {Observation of a new boson
  at a mass of 125 GeV with the CMS experiment at the LHC}},  {\em Phys.Lett.}
  {\bf B716} (2012) 30--61, [\href{http://xxx.lanl.gov/abs/1207.7235}{{\tt
  arXiv:1207.7235}}].

\bibitem{:2012afa}
{\bf ATLAS} Collaboration, G.~Aad et~al., {\it {Search for diphoton events with
  large missing transverse momentum in 7 TeV proton-proton collision data with
  the ATLAS detector}},  {\em Phys.Lett.} {\bf B718} (2012) 411--430,
  [\href{http://xxx.lanl.gov/abs/1209.0753}{{\tt arXiv:1209.0753}}].

\bibitem{Aad:2012cy}
{\bf ATLAS} Collaboration, G.~Aad et~al., {\it {Search for Extra Dimensions in
  diphoton events using proton-proton collisions recorded at $\sqrt{s}=7$ TeV
  with the ATLAS detector at the LHC}},
  \href{http://xxx.lanl.gov/abs/1210.8389}{{\tt arXiv:1210.8389}}.

\bibitem{CMS:2012un}
{\bf CMS} Collaboration, S.~Chatrchyan et~al., {\it {Search for supersymmetry
  in events with photons and low missing transverse energy in $pp$ collisions
  at $\sqrt{s}=7$ TeV}},  {\em Phys. Lett. B.} (2012)
  [\href{http://xxx.lanl.gov/abs/1210.2052}{{\tt arXiv:1210.2052}}].

\bibitem{:2012mx}
{\bf CMS} Collaboration, S.~Chatrchyan et~al., {\it {Search for new physics in
  events with photons, jets, and missing transverse energy in $pp$ collisions
  at $\sqrt{s}=7$ TeV}},  \href{http://xxx.lanl.gov/abs/1211.4784}{{\tt
  arXiv:1211.4784}}.

\bibitem{Binoth:1999qq}
T.~Binoth, J.~P. Guillet, E.~Pilon, and M.~Werlen, {\it {A Full next-to-leading
  order study of direct photon pair production in hadronic collisions}},  {\em
  Eur. Phys. J.} {\bf C16} (2000) 311--330,
  [\href{http://xxx.lanl.gov/abs/arXiv hep-ph/9911340}{{\tt arXiv
  hep-ph/9911340}}].

\bibitem{Bern:2002jx}
Z.~Bern, L.~J. Dixon, and C.~Schmidt, {\it {Isolating a light Higgs boson from
  the diphoton background at the CERN LHC}},  {\em Phys.Rev.} {\bf D66} (2002)
  074018, [\href{http://xxx.lanl.gov/abs/hep-ph/0206194}{{\tt
  hep-ph/0206194}}].

\bibitem{Balazs:1999yf}
C.~Balazs, P.~M. Nadolsky, C.~Schmidt, and C.~Yuan, {\it {Diphoton background
  to Higgs boson production at the LHC with soft gluon effects}},  {\em
  Phys.Lett.} {\bf B489} (2000) 157--162,
  [\href{http://xxx.lanl.gov/abs/hep-ph/9905551}{{\tt hep-ph/9905551}}].

\bibitem{Balazs:2006cc}
C.~Balazs, E.~L. Berger, P.~M. Nadolsky, and C.-P. Yuan, {\it {All-orders
  resummation for diphoton production at hadron colliders}},  {\em Phys.Lett.}
  {\bf B637} (2006) 235--240,
  [\href{http://xxx.lanl.gov/abs/hep-ph/0603037}{{\tt hep-ph/0603037}}].

\bibitem{Catani:2011qz}
S.~Catani, L.~Cieri, D.~de~Florian, G.~Ferrera, and M.~Grazzini, {\it {Diphoton
  production at hadron colliders: a fully-differential QCD calculation at
  NNLO}},  {\em Phys.Rev.Lett.} {\bf 108} (2012) 072001,
  [\href{http://xxx.lanl.gov/abs/1110.2375}{{\tt arXiv:1110.2375}}].

\bibitem{Hoeche:2009xc}
S.~Hoeche, S.~Schumann, and F.~Siegert, {\it {Hard photon production and
  matrix-element parton-shower merging}},  {\em Phys. Rev.} {\bf D81} (2010)
  034026, [\href{http://xxx.lanl.gov/abs/arXiv hep-ph/0912.3501}{{\tt arXiv
  hep-ph/0912.3501}}].

\bibitem{D'Errico:2011sd}
L.~D'Errico and P.~Richardson, {\it {Next-to-Leading-Order Monte Carlo
  Simulation of Diphoton Production in Hadronic Collisions}},  {\em JHEP} {\bf
  1202} (2012) 130, [\href{http://xxx.lanl.gov/abs/1106.3939}{{\tt
  arXiv:1106.3939}}].

\bibitem{Odaka:2012ry}
S.~Odaka and Y.~Kurihara, {\it {Consistent simulation of non-resonant diphoton
  production at hadron collisions with a custom-made parton shower}},  {\em
  Phys.Rev.} {\bf D85} (2012) 114022,
  [\href{http://xxx.lanl.gov/abs/1203.4038}{{\tt arXiv:1203.4038}}].

\bibitem{Glover:1993xc}
E.~N. Glover and A.~Morgan, {\it {Measuring the photon fragmentation function
  at LEP}},  {\em Z.Phys.} {\bf C62} (1994) 311--322.

\bibitem{GehrmannDeRidder:1997wx}
A.~Gehrmann-De~Ridder, T.~Gehrmann, and E.~N. Glover, {\it {Radiative
  corrections to the photon + 1 jet rate at LEP}},  {\em Phys.Lett.} {\bf B414}
  (1997) 354--361, [\href{http://xxx.lanl.gov/abs/hep-ph/9705305}{{\tt
  hep-ph/9705305}}].

\bibitem{GehrmannDeRidder:1998ba}
A.~Gehrmann-De~Ridder and E.~N. Glover, {\it {Final state photon production at
  LEP}},  {\em Eur.Phys.J.} {\bf C7} (1999) 29--48,
  [\href{http://xxx.lanl.gov/abs/hep-ph/9806316}{{\tt hep-ph/9806316}}].

\bibitem{DelDuca:2003uz}
V.~Del~Duca, F.~Maltoni, Z.~Nagy, and Z.~Trocsanyi, {\it {QCD radiative
  corrections to prompt diphoton production in association with a jet at hadron
  colliders}},  {\em JHEP} {\bf 0304} (2003) 059,
  [\href{http://xxx.lanl.gov/abs/hep-ph/0303012}{{\tt hep-ph/0303012}}].

\bibitem{Bern:2011pa}
Z.~Bern, G.~Diana, L.~Dixon, F.~Febres~Cordero, S.~Hoche, et~al., {\it {Driving
  Missing Data at Next-to-Leading Order}},  {\em Phys.Rev.} {\bf D84} (2011)
  114002, [\href{http://xxx.lanl.gov/abs/1106.1423}{{\tt arXiv:1106.1423}}].

\bibitem{Jager:2010aj}
B.~Jager, {\it {Next-to-leading order QCD corrections to photon production via
  weak-boson fusion}},  {\em Phys.Rev.} {\bf D81} (2010) 114016,
  [\href{http://xxx.lanl.gov/abs/1004.0825}{{\tt arXiv:1004.0825}}].

\bibitem{Cullen:2011ac}
G.~Cullen, N.~Greiner, G.~Heinrich, G.~Luisoni, P.~Mastrolia, et~al., {\it
  {Automated One-Loop Calculations with GoSam}},  {\em Eur.Phys.J.} {\bf C72}
  (2012) 1889, [\href{http://xxx.lanl.gov/abs/1111.2034}{{\tt
  arXiv:1111.2034}}].

\bibitem{Salam:2009jx}
G.~P. Salam, {\it {Towards Jetography}},  {\em Eur.Phys.J.} {\bf C67} (2010)
  637--686, [\href{http://xxx.lanl.gov/abs/0906.1833}{{\tt arXiv:0906.1833}}].

\bibitem{Frixione:1998hn}
S.~Frixione, {\it {Isolated photons in perturbative QCD}},  {\em Phys.Lett.}
  {\bf B429} (1998) 369--374, [\href{http://xxx.lanl.gov/abs/arXiv
  hep-ph/9801442}{{\tt arXiv hep-ph/9801442}}].

\bibitem{AlcarazMaestre:2012vp}
J.~Alcaraz~Maestre et~al., {\it {The Les Houches 2011 NLO Multileg and MC
  Working Groups: Summary Report}},  {\em Proceedings} (2012)
  [\href{http://xxx.lanl.gov/abs/1203.6803}{{\tt arXiv:1203.6803}}].

\bibitem{Kunszt:1992np}
Z.~Kunszt and Z.~Trocsanyi, {\it {QCD corrections to photon production in
  association with hadrons in e+ e- annihilation}},  {\em Nucl.Phys.} {\bf
  B394} (1993) 139--168, [\href{http://xxx.lanl.gov/abs/hep-ph/9207232}{{\tt
  hep-ph/9207232}}].

\bibitem{Ackerstaff:1997nha}
{\bf OPAL} Collaboration, K.~Ackerstaff et~al., {\it {Measurement of the quark
  to photon fragmentation function through the inclusive production of prompt
  photons in hadronic Z0 decays}},  {\em Eur.Phys.J.} {\bf C2} (1998) 39--48,
  [\href{http://xxx.lanl.gov/abs/hep-ex/9708020}{{\tt hep-ex/9708020}}].

\bibitem{GehrmannDeRidder:2006wz}
A.~Gehrmann-De~Ridder, T.~Gehrmann, and E.~Poulsen, {\it {Isolated photons in
  deep inelastic scattering}},  {\em Phys.Rev.Lett.} {\bf 96} (2006) 132002,
  [\href{http://xxx.lanl.gov/abs/hep-ph/0601073}{{\tt hep-ph/0601073}}].

\bibitem{Chekanov:2004wr}
{\bf ZEUS} Collaboration, S.~Chekanov et~al., {\it {Observation of isolated
  high E(T) photons in deep inelastic scattering}},  {\em Phys.Lett.} {\bf
  B595} (2004) 86--100, [\href{http://xxx.lanl.gov/abs/hep-ex/0402019}{{\tt
  hep-ex/0402019}}].

\bibitem{Chekanov:2009dq}
{\bf ZEUS} Collaboration, S.~Chekanov et~al., {\it {Measurement of isolated
  photon production in deep inelastic ep scattering}},  {\em Phys.Lett.} {\bf
  B687} (2010) 16--25, [\href{http://xxx.lanl.gov/abs/0909.4223}{{\tt
  arXiv:0909.4223}}].

\bibitem{Aaron:2007aa}
{\bf H1} Collaboration, F.~Aaron et~al., {\it {Measurement of isolated photon
  production in deep-inelastic scattering at HERA}},  {\em Eur.Phys.J.} {\bf
  C54} (2008) 371--387, [\href{http://xxx.lanl.gov/abs/0711.4578}{{\tt
  arXiv:0711.4578}}].

\bibitem{Buskulic:1995au}
{\bf ALEPH} Collaboration, D.~Buskulic et~al., {\it {First measurement of the
  quark to photon fragmentation function}},  {\em Z.Phys.} {\bf C69} (1996)
  365--378.

\bibitem{GehrmannDeRidder:1997gf}
A.~Gehrmann-De~Ridder and E.~N. Glover, {\it {A Complete ${\cal
  O}(\alpha\alpha_s)$ calculation of the photon + 1 jet rate in $e^+e^-$
  annihilation}},  {\em Nucl.Phys.} {\bf B517} (1998) 269--323,
  [\href{http://xxx.lanl.gov/abs/hep-ph/9707224}{{\tt hep-ph/9707224}}].

\bibitem{GehrmannDeRidder:2006vn}
A.~Gehrmann-De~Ridder, T.~Gehrmann, and E.~Poulsen, {\it {Measuring the Photon
  Fragmentation Function at HERA}},  {\em Eur.Phys.J.} {\bf C47} (2006)
  395--411, [\href{http://xxx.lanl.gov/abs/hep-ph/0604030}{{\tt
  hep-ph/0604030}}].

\bibitem{Altarelli:1977zs}
G.~Altarelli and G.~Parisi, {\it {Asymptotic Freedom in Parton Language}},
  {\em Nucl.Phys.} {\bf B126} (1977) 298.

\bibitem{Owens:1986mp}
J.~Owens, {\it {Large Momentum Transfer Production of Direct Photons, Jets, and
  Particles}},  {\em Rev.Mod.Phys.} {\bf 59} (1987) 465.

\bibitem{Gluck:1992zx}
M.~Gluck, E.~Reya, and A.~Vogt, {\it {Parton fragmentation into photons beyond
  the leading order}},  {\em Phys.Rev.} {\bf D48} (1993) 116.

\bibitem{Bourhis:1997yu}
L.~Bourhis, M.~Fontannaz, and J.~Guillet, {\it {Quarks and gluon fragmentation
  functions into photons}},  {\em Eur.Phys.J.} {\bf C2} (1998) 529--537,
  [\href{http://xxx.lanl.gov/abs/arXiv hep-ph/9704447}{{\tt arXiv
  hep-ph/9704447}}].

\bibitem{Denner:2009gx}
A.~Denner, S.~Dittmaier, T.~Gehrmann, and C.~Kurz, {\it {Electroweak
  corrections to three-jet production in electron-positron annihilation}},
  {\em Phys.Lett.} {\bf B679} (2009) 219--222,
  [\href{http://xxx.lanl.gov/abs/0906.0372}{{\tt arXiv:0906.0372}}].

\bibitem{Denner:2010ia}
A.~Denner, S.~Dittmaier, T.~Gehrmann, and C.~Kurz, {\it {Electroweak
  corrections to hadronic event shapes and jet production in e+e-
  annihilation}},  {\em Nucl.Phys.} {\bf B836} (2010) 37--90,
  [\href{http://xxx.lanl.gov/abs/1003.0986}{{\tt arXiv:1003.0986}}].

\bibitem{Denner:2009gj}
A.~Denner, S.~Dittmaier, T.~Kasprzik, and A.~Muck, {\it {Electroweak
  corrections to W + jet hadroproduction including leptonic W-boson decays}},
  {\em JHEP} {\bf 0908} (2009) 075,
  [\href{http://xxx.lanl.gov/abs/0906.1656}{{\tt arXiv:0906.1656}}].

\bibitem{Denner:2011vu}
A.~Denner, S.~Dittmaier, T.~Kasprzik, and A.~Muck, {\it {Electroweak
  corrections to dilepton + jet production at hadron colliders}},  {\em JHEP}
  {\bf 1106} (2011) 069, [\href{http://xxx.lanl.gov/abs/1103.0914}{{\tt
  arXiv:1103.0914}}].

\bibitem{Catani:2002ny}
S.~Catani, M.~Fontannaz, J.~Guillet, and E.~Pilon, {\it {Cross-section of
  isolated prompt photons in hadron hadron collisions}},  {\em JHEP} {\bf 0205}
  (2002) 028, [\href{http://xxx.lanl.gov/abs/arXiv hep-ph/0204023}{{\tt arXiv
  hep-ph/0204023}}].

\bibitem{Belghobsi:2009hx}
Z.~Belghobsi, M.~Fontannaz, J.-P. Guillet, G.~Heinrich, E.~Pilon, et~al., {\it
  {Photon - Jet Correlations and Constraints on Fragmentation Functions}},
  {\em Phys.Rev.} {\bf D79} (2009) 114024,
  [\href{http://xxx.lanl.gov/abs/0903.4834}{{\tt arXiv:0903.4834}}].

\bibitem{Stelzer:1994ta}
T.~Stelzer and W.~Long, {\it {Automatic generation of tree level helicity
  amplitudes}},  {\em Comput.Phys.Commun.} {\bf 81} (1994) 357--371,
  [\href{http://xxx.lanl.gov/abs/hep-ph/9401258}{{\tt hep-ph/9401258}}].

\bibitem{Alwall:2007st}
J.~Alwall, P.~Demin, S.~de~Visscher, R.~Frederix, M.~Herquet, et~al., {\it
  {MadGraph/MadEvent v4: The New Web Generation}},  {\em JHEP} {\bf 0709}
  (2007) 028, [\href{http://xxx.lanl.gov/abs/0706.2334}{{\tt
  arXiv:0706.2334}}].

\bibitem{Frederix:2008hu}
R.~Frederix, T.~Gehrmann, and N.~Greiner, {\it {Automation of the Dipole
  Subtraction Method in MadGraph/MadEvent}},  {\em JHEP} {\bf 0809} (2008) 122,
  [\href{http://xxx.lanl.gov/abs/0808.2128}{{\tt arXiv:0808.2128}}].

\bibitem{Frederix:2010cj}
R.~Frederix, T.~Gehrmann, and N.~Greiner, {\it {Integrated dipoles with
  MadDipole in the MadGraph framework}},  {\em JHEP} {\bf 1006} (2010) 086,
  [\href{http://xxx.lanl.gov/abs/1004.2905}{{\tt arXiv:1004.2905}}].

\bibitem{Catani:1996vz}
S.~Catani and M.~Seymour, {\it {A General algorithm for calculating jet
  cross-sections in NLO QCD}},  {\em Nucl.Phys.} {\bf B485} (1997) 291--419,
  [\href{http://xxx.lanl.gov/abs/hep-ph/9605323}{{\tt hep-ph/9605323}}].

\bibitem{Maltoni:2002qb}
F.~Maltoni and T.~Stelzer, {\it {MadEvent: Automatic event generation with
  MadGraph}},  {\em JHEP} {\bf 0302} (2003) 027,
  [\href{http://xxx.lanl.gov/abs/hep-ph/0208156}{{\tt hep-ph/0208156}}].

\bibitem{Nogueira:1991ex}
P.~Nogueira, {\it {Automatic Feynman graph generation}},  {\em J.Comput.Phys.}
  {\bf 105} (1993) 279--289.

\bibitem{Vermaseren:2000nd}
J.~Vermaseren, {\it {New features of FORM}},
  \href{http://xxx.lanl.gov/abs/math-ph/0010025}{{\tt math-ph/0010025}}.

\bibitem{Kuipers:2012rf}
J.~Kuipers, T.~Ueda, J.~Vermaseren, and J.~Vollinga, {\it {FORM version 4.0}},
  \href{http://xxx.lanl.gov/abs/1203.6543}{{\tt arXiv:1203.6543}}.

\bibitem{Cullen:2010jv}
G.~Cullen, M.~Koch-Janusz, and T.~Reiter, {\it {Spinney: A Form Library for
  Helicity Spinors}},  {\em Comput.Phys.Commun.} {\bf 182} (2011) 2368--2387,
  [\href{http://xxx.lanl.gov/abs/1008.0803}{{\tt arXiv:1008.0803}}].

\bibitem{Reiter:2009ts}
T.~Reiter, {\it {Optimising Code Generation with haggies}},  {\em
  Comput.Phys.Commun.} {\bf 181} (2010) 1301--1331,
  [\href{http://xxx.lanl.gov/abs/0907.3714}{{\tt arXiv:0907.3714}}].

\bibitem{Ossola:2006us}
G.~Ossola, C.~G. Papadopoulos, and R.~Pittau, {\it {Reducing full one-loop
  amplitudes to scalar integrals at the integrand level}},  {\em Nucl.Phys.}
  {\bf B763} (2007) 147--169,
  [\href{http://xxx.lanl.gov/abs/hep-ph/0609007}{{\tt hep-ph/0609007}}].

\bibitem{Ellis:2007br}
R.~Ellis, W.~Giele, and Z.~Kunszt, {\it {A Numerical Unitarity Formalism for
  Evaluating One-Loop Amplitudes}},  {\em JHEP} {\bf 0803} (2008) 003,
  [\href{http://xxx.lanl.gov/abs/0708.2398}{{\tt arXiv:0708.2398}}].

\bibitem{Mastrolia:2008jb}
P.~Mastrolia, G.~Ossola, C.~Papadopoulos, and R.~Pittau, {\it {Optimizing the
  Reduction of One-Loop Amplitudes}},  {\em JHEP} {\bf 0806} (2008) 030,
  [\href{http://xxx.lanl.gov/abs/0803.3964}{{\tt arXiv:0803.3964}}].

\bibitem{Mastrolia:2010nb}
P.~Mastrolia, G.~Ossola, T.~Reiter, and F.~Tramontano, {\it {Scattering
  AMplitudes from Unitarity-based Reduction Algorithm at the Integrand-level}},
   {\em JHEP} {\bf 1008} (2010) 080,
  [\href{http://xxx.lanl.gov/abs/1006.0710}{{\tt arXiv:1006.0710}}].

\bibitem{Heinrich:2010ax}
G.~Heinrich, G.~Ossola, T.~Reiter, and F.~Tramontano, {\it {Tensorial
  Reconstruction at the Integrand Level}},  {\em JHEP} {\bf 1010} (2010) 105,
  [\href{http://xxx.lanl.gov/abs/1008.2441}{{\tt arXiv:1008.2441}}].

\bibitem{Binoth:2008uq}
T.~Binoth, J.-P. Guillet, G.~Heinrich, E.~Pilon, and T.~Reiter, {\it {Golem95:
  A Numerical program to calculate one-loop tensor integrals with up to six
  external legs}},  {\em Comput.Phys.Commun.} {\bf 180} (2009) 2317--2330,
  [\href{http://xxx.lanl.gov/abs/0810.0992}{{\tt arXiv:0810.0992}}].

\bibitem{Cullen:2011kv}
G.~Cullen, J.~Guillet, G.~Heinrich, T.~Kleinschmidt, E.~Pilon, et~al., {\it
  {Golem95C: A library for one-loop integrals with complex masses}},  {\em
  Comput.Phys.Commun.} {\bf 182} (2011) 2276--2284,
  [\href{http://xxx.lanl.gov/abs/1101.5595}{{\tt arXiv:1101.5595}}].

\bibitem{Dittmaier:1999mb}
S.~Dittmaier, {\it {A General approach to photon radiation off fermions}},
  {\em Nucl.Phys.} {\bf B565} (2000) 69--122,
  [\href{http://xxx.lanl.gov/abs/hep-ph/9904440}{{\tt hep-ph/9904440}}].

\bibitem{Gehrmann:2010ry}
T.~Gehrmann and N.~Greiner, {\it {Photon Radiation with MadDipole}},  {\em
  JHEP} {\bf 1012} (2010) 050, [\href{http://xxx.lanl.gov/abs/1011.0321}{{\tt
  arXiv:1011.0321}}].

\bibitem{Nagy:1998bb}
Z.~Nagy and Z.~Trocsanyi, {\it {Next-to-leading order calculation of four jet
  observables in electron positron annihilation}},  {\em Phys.Rev.} {\bf D59}
  (1999) 014020, [\href{http://xxx.lanl.gov/abs/hep-ph/9806317}{{\tt
  hep-ph/9806317}}].

\bibitem{Cacciari:2008gp}
M.~Cacciari, G.~P. Salam, and G.~Soyez, {\it {The Anti-k(t) jet clustering
  algorithm}},  {\em JHEP} {\bf 0804} (2008) 063,
  [\href{http://xxx.lanl.gov/abs/0802.1189}{{\tt arXiv:0802.1189}}].

\bibitem{Cacciari:2011ma}
M.~Cacciari, G.~P. Salam, and G.~Soyez, {\it {FastJet User Manual}},  {\em
  Eur.Phys.J.} {\bf C72} (2012) 1896,
  [\href{http://xxx.lanl.gov/abs/1111.6097}{{\tt arXiv:1111.6097}}].

\bibitem{Cacciari:2005hq}
M.~Cacciari and G.~P. Salam, {\it {Dispelling the $N^{3}$ myth for the $k_t$
  jet-finder}},  {\em Phys.Lett.} {\bf B641} (2006) 57--61,
  [\href{http://xxx.lanl.gov/abs/hep-ph/0512210}{{\tt hep-ph/0512210}}].

\bibitem{Ball:2012cx}
R.~D. Ball, V.~Bertone, S.~Carrazza, C.~S. Deans, L.~Del~Debbio, et~al., {\it
  {Parton distributions with LHC data}},  {\em Nucl.Phys.} {\bf B867} (2013)
  244--289, [\href{http://xxx.lanl.gov/abs/1207.1303}{{\tt arXiv:1207.1303}}].

\bibitem{Binoth:2009wk}
T.~Binoth, T.~Gleisberg, S.~Karg, N.~Kauer, and G.~Sanguinetti, {\it {NLO QCD
  corrections to ZZ+ jet production at hadron colliders}},  {\em Phys.Lett.}
  {\bf B683} (2010) 154--159, [\href{http://xxx.lanl.gov/abs/0911.3181}{{\tt
  arXiv:0911.3181}}].

\bibitem{Campbell:2014yka}
  J.~M.~Campbell and C.~Williams,
  {\it Triphoton production at hadron colliders},
 [\href{http://xxx.lanl.gov/abs/1403.2641}{{\tt arXiv:1403.2641}}].
  
\end{thebibliography}
\providecommand{\href}[2]{#2}\begingroup\raggedright\endgroup

\end{document}